\documentclass[12pt]{article}
\DeclareMathAlphabet{\scr}{U}{rsfs}{m}{n}
\usepackage{latexsym,url}
\usepackage[mathscr]{eucal}
\usepackage{amsfonts}
\usepackage{amscd}
\usepackage{cite}         
\usepackage{amssymb}
\usepackage{color}
\usepackage{colordvi}
\usepackage[centertags]{amsmath}
\usepackage{enumerate}
\usepackage{graphicx}
\usepackage{booktabs}
\usepackage{bbm}
\usepackage{bbold}
\usepackage{dsfont}
\usepackage{slashed}

\newcommand{\cleqn}{\setcounter{equation}{0}}
\newcommand{\newc}{\newcommand}
\newc{\be}{\begin{equation}}
\newc{\ee}{\end{equation}}
\newc{\bea}{\begin{eqnarray}}
\newc{\eea}{\end{eqnarray}}
\newc{\ben}{\begin{equation*}}
\newc{\een}{\end{equation*}}
\newc{\bean}{\begin{eqnarray*}}
\newc{\eean}{\end{eqnarray*}}
\newc{\ol}{\overline}
\newc{\wt}{\widetilde}
\newc{\bs}{\boldsymbol}
\newc{\m}{\mathcal}
\newc{\la}{\lambda}
\newc{\lra}{\longrightarrow}
\newc{\vp}{\varphi}
\newc{\ti}{\tilde}
\def\nn{\nonumber}
\newc{\vev}[1]{\langle#1 \rangle}
 
\newcommand{\id}{\mathds 1}
\newcommand{\re}[1]{\mathbf{#1}}
\newcommand{\rb}[1]{\mathbf{\overline{#1}}}

\begin{document}

\title{\hfill ~\\[-30mm]
          \hfill\mbox{\small  QFET-2016-16}\\[-3.5mm]
          \hfill\mbox{\small  SI-HEP-2016-26}\\[18mm]
       \textbf{
Lepton-flavour violation in a Pati-Salam\\[2mm]
model with gauged flavour symmetry\\[4mm]
}}

\author{
Thorsten Feldmann\,\footnote{E-mail: {\tt thorsten.feldmann@uni-siegen.de}}
,~~
Christoph Luhn\,\footnote{E-mail: {\tt christoph.luhn@uni-siegen.de}}
,~~
Paul Moch\,\footnote{E-mail: {\tt paulmoch@physik.uni-siegen.de}}\\[8mm]
\it{\small Theoretische Physik 1, Naturwissenschaftlich-Technische
    Fakult\"at, Universit\"at Siegen,}\\
\it{\small Walter-Flex-Stra{\ss}e 3, 57068 Siegen, Germany}
}

\date{}

\maketitle

\begin{abstract}
\noindent

Combining Pati-Salam (PS) and flavour symmetries in a renormalisable setup, 
we devise a scenario which produces realistic masses for the charged
leptons. Flavour-symmetry breaking scalar fields in the adjoint
representations of the PS gauge group are responsible for generating different
flavour structures for up- and down-type quarks as well as for leptons.
The model is characterised by new heavy fermions which mix with the Standard
Model quarks and leptons. In particular, the partners for the third fermion
generation induce sizeable sources of flavour violation. Focusing on the
charged-lepton sector, we scrutinise the model with respect to its
implications for lepton-flavour violating processes such as $\mu \rightarrow
e\gamma$,  $\mu\rightarrow 3e$ and muon conversion in nuclei.

\end{abstract}
\thispagestyle{empty}
\vfill

\newpage \setcounter{page}{1}




\section{Introduction\label{sec:1}}
\cleqn

The successful description of gauge interactions is arguably one of the
most attractive features of the Standard Model (SM) of particle physics. Their
structure is dictated by the symmetry $SU(3)\times SU(2)\times U(1)$,
where each factor comes with its own gauge coupling constant. Adding to this the
electroweak symmetry breaking sector introduces two extra parameters, the
vacuum expectation value (VEV) and the mass of the Higgs boson, discovered at
the LHC in 2012~\cite{Aad:2012tfa,Chatrchyan:2012xdj}. 
Unlike the gauge interactions, the structure of the Yukawa
sector, which provides the seed of quark and lepton masses and mixing, is much
less understood. Ever since I.~I.~Rabi phrased his old question about the
muon, ``Who ordered it?'', physicists have been trying hard to unravel the
origin of flavour. 

Strictly speaking, the flavour structure of the SM does not require an
underlying principle. Assigning appropriate numerical values to the a priori
undetermined Yukawa coupling constants is sufficient to consistently
parameterise flavour in the SM. Yet, when considering extensions of the SM,
the need for a theory of flavour becomes more pressing. In general, the
implications of theories beyond the SM for low-energy physics can be
formulated using the SM effective field theory approach, with a total of 2499
possible operators at dimension six~\cite{Buchmuller:1985jz,Grzadkowski:2010es,Alonso:2013hga}. 
Many of these non-renormalisable operators entail new sources of flavour and
CP violation. Without a mechanism that controls the size and the structure of
these new couplings, the effective theory would generally not be compatible
with experimental results for certain flavour
observables~\cite{Altmannshofer:2015zpk}. 

Although not a theory of flavour, the concept of minimal flavour violation
(MFV) provides an appealing framework for constructing higher-dimensional
operators which efficiently suppresses flavour changing processes beyond the
SM~\cite{Chivukula:1987py,Hall:1990ac,Buras:2000dm,D'Ambrosio:2002ex,Cirigliano:2005ck}. 
The idea is based on a symmetry principle, more precisely on the maximal
flavour symmetry of the SM in the absence of the Yukawa couplings. As each of
the five fermionic SM multiplets comes in three generations, it is given by $U(3)^5$.
Furthermore, MFV postulates that the only flavour symmetry breaking entities
are the Yukawa matrices themselves, whose occurrence in the higher-dimensional
operators of an effective field theory is controlled by the original flavour symmetry.

In order to embed the concept of MFV into a high-energy theory, it is
necessary to promote the Yukawa matrices to scalar fields. In such a
setup, the SM Yukawa couplings originate dynamically from non-renormalisable
operators after inserting VEVs for the matrix-valued scalar flavon fields. 
Sequential flavour symmetry breaking~\cite{Feldmann:2009dc} can be realised by
hierarchical vacuum configurations which are derived from appropriate flavour
symmetric potentials as discussed
e.g.~in~\cite{Alonso:2011yg,Nardi:2011st,Alonso:2012fy,Espinosa:2012uu,Alonso:2013mca,Alonso:2013nca,Alonso:2013dba,Fong:2013dnk,Merlo:2015nra}. Massless 
Goldstone modes are avoided by gauging the flavour symmetry. Pursuing this
idea in a renormalisable setup, Grinstein, Redi and Villadoro (GRV) proposed a
model~\cite{Grinstein:2010ve} in which new heavy partners of the SM fermions
mediate the coupling of left- and right-chiral quarks and leptons to the Higgs, with the
resulting effective Yukawa matrices being inversely proportional to the VEVs of
the matrix-valued flavon fields. The extension of the fermionic particle
content was chosen such as to cancel all gauge anomalies of the SM and flavour
symmetries (see also~\cite{Albrecht:2010xh}).  

The basic idea of GRV has been extended and applied in the context of the
SM~\cite{D'Agnolo:2012ie,Bishara:2015mha},
grand unified theories
(GUTs)~\cite{Feldmann:2010yp,Guadagnoli:2011id,Mohapatra:2012km,Feldmann:2015zwa}, 
and supersymmetric theories~\cite{Krnjaic:2012aj,Franceschini:2013ne}. The
combination of flavour and GUT symmetries is of particular interest as it
allows to unify the theory both horizontally (by organising the three
generations of fermions into triplets) as well as vertically (by combining
independent SM multiplets into a single GUT multiplet). For reviews, see e.g.~\cite{Chen:2003zv,King:2003jb,Mohapatra:2006gs,Babu:2009fd,Altarelli:2010gt,King:2013eh,King:2014nza,King:2015aea}.
Among the various GUT symmetries, the Pati-Salam (PS) gauge
symmetry~\cite{Pati:1973uk,Pati:1974yy} provides 
a rich playground in which gauge coupling unification can be realised via
several intermediate mass scales~\cite{Kilian:2006hh,Howl:2007hq,Calibbi:2009cp,Braam:2009fi,DeRomeri:2011ie,Arbelaez:2013hr,Arbelaez:2013nga,Hartmann:2014fya}. 
Flavour models which are compatible with an underlying PS gauge
group have been constructed abundantly in the literature, see e.g.~\cite{King:2003rf,King:2006me,King:2006np,King:2009mk,Dutta:2009bj,King:2009tj,Toorop:2010yh,BhupalDev:2011gi,deMedeirosVarzielas:2011wx,BhupalDev:2012nm,Hartmann:2014ppa,King:2014iia}.
A concrete setup which embeds the GRV mechanism in an explicit left-right
(i.e.~$Z_2$) symmetric PS GUT was put forth in~\cite{Feldmann:2015zwa}. With
all SM multiplets (plus the right-chiral neutrino) being unified into only two PS
multiplets, the full symmetry of the Lagrangian is given by
\be \nonumber
\underbrace{\Big( SU(4)\times SU(2)\times SU(2)' \Big)}_{\text{Pati-Salam}}\;\times\; 
\underbrace{\Big( SU(3)_{I}\times SU(3)_{II} \Big)}_{\text{flavour}}
    \;\times\; Z_2\ ,
\ee
where the $Z_2$ maps the multiplet 
$(\omega_c,\omega,\omega')(\omega_{I},\omega_{II}) $ into
$(\overline{\omega}_c,\omega',\omega)
(\overline{\omega}_{II},\overline{\omega}_{I})$. 
As in~\cite{Feldmann:2015zwa}, we assume that the $U(1)$ factors within
$U(3)_I \times U(3)_{II}$ are explicitly broken in the scalar potential,
e.g. by terms involving the determinant of the matrix-valued flavons.

While the phenomenology of the quark sector was investigated thoroughly for 
this setup, the study in~\cite{Feldmann:2015zwa} stopped short of a similar
analysis of the lepton sector. It is the purpose of the present article to complete
this work by formulating a viable extension of the lepton sector, involving
additional Pati-Salam and flavour-symmetry breaking scalar flavon fields, and
studying the expected signatures for lepton-flavour violating (LFV) processes. 

The layout of the remainder of this paper is as follows. In
Section~\ref{sec:2}, we recapitulate the main ingredients of the model
in~\cite{Feldmann:2015zwa} and present the extension of the scalar
sector necessary to generate viable charged-lepton masses. 
Section~\ref{sec:3} discusses the diagonalisation of the charged- and
neutral-lepton mass matrices, as well as the effect this change of basis has
on the gauge-kinetic terms. We explicitly state the resulting anomalous
couplings of the leptons to the electroweak gauge bosons $W$ and $Z$, briefly
also commenting on the anomalous Higgs coupling. 
In Section~\ref{sec:4}, we relate these anomalous couplings to the LFV
observables 
$\mu \rightarrow e\gamma$,  
$\mu\rightarrow 3e$ 
and muon conversion in the vicinity of nuclei.
The phenomenological implications of our model as derived from a numerical
parameter scan are discussed in Section~\ref{sec:5}.
We conclude in Section~\ref{sec:6}.
Appendix~\ref{app:basis-transformations} provides technical details on the 
transformation from the flavour basis to the mass basis.




\section{A Pati-Salam model with viable charged leptons\label{sec:2}}
\cleqn

The setup of the left-right symmetric Pati-Salam model with gauged $SU(3)_{I}
\times SU(3)_{II}$ flavour symmetry has been discussed 
in~\cite{Feldmann:2015zwa}. In order to make the present paper self-contained,
we begin with a brief recapitulation of the main ideas and the required
ingredients of the model. The left- and right-chiral Pati-Salam multiplets
$q_{L,R}$ contain all SM fermions. Additional fermionic partners
($\Sigma_{L,R}$, $\Xi_{L,R}$) are introduced in order to formulate a
renormalisable model in which the flavour symmetry is broken by VEVs of
matrix-valued scalar fields $S$ and $T'$. These
flavon fields transform trivially under PS, except for $T'$ which
furnishes a triplet representation of the PS gauge factor $SU(2)'$. The
combination of flavons in both the singlet and triplet representation of
$SU(2)'$ is necessary to distinguish up-type from down-type flavour
structures. To first approximation, the Yukawa matrices of the light (i.e.~SM)
fermions are obtained by integrating out the heavy partners. More accurately,
it is necessary to diagonalise a $9\times 9$ mass matrix in order to
identify the correct mass eigenstates. This is particularly relevant for the
up-type sector where the two partners of the top quark are only marginally
heavier than the top itself. 

As with every gauge symmetry, anomaly considerations put severe constraints on
the fermionic particle content of the model. The resulting need for introducing
further fermions in our PS setup can be naturally combined with the construction of a
realistic (Majorana) neutrino sector. To this end, we introduce PS neutral
fermions $\Theta_{L,R}$ which acquire Majorana masses through extra scalars
$S^{(\prime)}_\nu$. As a side effect, these additional flavour symmetry
breaking scalar fields are responsible for decoupling the flavour gauge bosons from
low-energy physics. The coupling of $\Theta_L$ to
the neutral component of $\Sigma_R$ induces a heavy Majorana mass for the
latter, which in turn generates light neutrino masses via the seesaw mechanism. 
We refer the reader to~\cite{Feldmann:2015zwa} for further details on the
construction of the model. 

The complete particle content of the Pati-Salam model as defined
in~\cite{Feldmann:2015zwa} is shown in Table~\ref{tab:fullfieldcontent}. It
yields a renormalisable Yukawa Lagrangian of the form
\be
\mathcal L_{\text{Yuk}}~=~ \mathcal L^q_{\text{Yuk}} \,+\,\mathcal
L^\nu_{\text{Yuk}} \ ,\label{eq:lagrangian}
\ee
with
  \begin{align}
\label{eq:fermionlagrangian}
     \mathcal{L}^q_\text{Yuk} \,= 
     &\phantom{+}\; \lambda\;\overline{q}_{L}\,H\,\Sigma_{R}
     + \overline{\Sigma}_L\,\left(\kappa_S\,S+\kappa_T\,T'\right)\,\Sigma_R
     + M \,\overline{\Sigma}_L\, q_{R} \;+\mathrm{h.c.} \cr 
     &+\lambda\;\overline{\Xi}_{L}\,H\,q_R
     + \overline{\Xi}_L\,\left(\kappa_S\,S+\kappa_T\,T\right)\,\Xi_R
     + M \,\overline{q}_L\,\Xi_{R} \;+\mathrm{h.c.} \ ,
\end{align}
and
\begin{align}
\label{eq:fermionlagrangian-nu}
     \mathcal{L}^\nu_\text{Yuk} \,\sim 
     &\phantom{+}\:\overline{\Theta}_{L}\,\Phi'\,\Sigma_{R}
     +\mbox{$\frac{1}{2}$} \, \overline{\Theta}_L\, \,S_\nu\,\overline\Theta_L \;+\mathrm{h.c.} \cr 
      &+\overline{\Xi}_{L}\,\Phi\,\Theta_{R}\hspace{0.25mm}
     + \mbox{$\frac{1}{2}$} \,{\Theta}_R\, \,S'_\nu\,\Theta_R \,+\mathrm{h.c.} \hspace{42.85mm} \cr 
     &+\overline{\Theta}_{L}\,S^\dagger \,\Theta_{R}  \;+\mathrm{h.c.} \ .
  \end{align}
The Lagrangian in Eq.~\eqref{eq:fermionlagrangian} describes the Yukawa
structure of the charged fermions. A comprehensive discussion of the quark
sector can be found in~\cite{Feldmann:2015zwa}. Turning to the charged
leptons, we observe that the effective Yukawa matrix $Y_\ell$ is identical to the
down-type quark Yukawa matrix $Y_d$. Although this provides a reasonable first
approximation, it is clear that an extension of some sort is required to
accommodate a fully realistic fermion mass pattern. We have already outlined
one such possibility in Appendix~A.2 of~\cite{Feldmann:2015zwa}, and it is the
purpose of this paper to work out the lepton-flavour phenomenology of such an
extension.

\begin{table}[t]
\centering
 \begin{tabular}{lccc}
   \toprule 
   ~~~~~~~ 
& Pati-Salam Symmetry & Flavour Symmetry & VEV\\  
& ~$SU(4)\times SU(2)\times SU(2)'$~ & $SU(3)_{I}\times SU(3)_{II}$ &  \\
   \midrule
   ~$\overline{q}_L$ & $(\rb4\,,\,\re2\,,\,\re1)$ & $(\rb3\,,\,\re1)$ &--- \\
   ~$q_R$ & $(\re4\,,\,\re1\,,\,\re2)$& $(\re1\,,\,\re3)$ &--- \\
   ~$H$  & $(\re1\,,\,\re2\,,\,\re2)$ & $(\re1\,,\,\re1)$ & $v_{u,d}$\\[1ex]
   ~$\overline{\Sigma}_L$ & $(\rb4\,,\,\re1\,,\,\re2)$ & $(\re1\,,\,\rb3)$ &--- \\
   ~$\Sigma_R$ & $(\re4\,,\,\re1\,,\,\re2)$ & $(\re3\,,\,\re1)$ &---  \\
   ~$\overline{\Xi}_L$ & $(\rb4\,,\,\re2\,,\,\re1)$ & $(\re1\,,\,\rb3)$ &--- \\
   ~$\Xi_R$ & $(\re4\,,\,\re2\,,\,\re1)$ & $(\re3\,,\,\re1)$ & ---\\[1ex]
   ~$T^{}_{1,15}$ & $(\re1+ \bf{15}\,,\,\re3\,,\,\re1)$ & $(\rb3\,,\,\re3)$ & 0 \\
   ~$T'_{1,15}$ & $(\re1+ \bf{15}\,,\,\re1\,,\,\re3)$ & $(\rb3\,,\,\re3)$ & $ t'_{1,15} \,M$ \\
   ~$S^{}_{1,15}$ & $(\re1 + \bf{15}\,,\,\re1\,,\,\re1)$ & $(\rb3\,,\,\re3)$ & $s^{}_{1,15}\,M$ \\\midrule
   ~$\overline\Theta_L$ & $(\re1\,,\,\re1\,,\,\re1)$ & $(\rb3\,,\,\re8)$ & ---\\
   ~$\Theta_R$ & $(\re1\,,\,\re1\,,\,\re1)$ & $(\re8\,,\,\re3)$ & ---\\[1ex]
   ~$S_\nu$ & $(\re1\,,\,\re1\,,\,\re1)$ & $(\re6\,,\,\re1)$ & $s_\nu \Lambda_\nu$\\
   ~$S'_\nu$ & $(\re1\,,\,\re1\,,\,\re1)$ & $(\re1\,,\,\rb6)$  & $s'_\nu \Lambda_\nu$ \\[1ex]
   ~$\Phi$ & $(\re4\,,\,\re2\,,\,\re1)$ & $(\re8\,,\,\re1)$ & 0\\
   ~$\Phi'$ & $(\rb4\,,\,\re1\,,\,\re2)$ & $(\re1\,,\,\re8)$ & $\varphi'\Lambda_\varphi$ \\
   \bottomrule
 \end{tabular}
\caption{The particle content of the Pati-Salam model with imposed flavour
  symmetry as defined in~\cite{Feldmann:2015zwa} and augmented by flavon
  fields in the adjoint representation of $SU(4)$. Left- and right-chiral
  fermions $\psi_{L,R}$ are denoted by subscripts $L$ and $R$,
  respectively. The VEVs of the scalar fields are given in the rightmost
  column. The lower part of the table shows fields necessary for generating
  Majorana neutrino masses.}
\label{tab:fullfieldcontent}
\end{table}

The idea consists in enlarging the scalar sector by flavour symmetry breaking
flavon fields which transform in the adjoint representation of $SU(4)$,
i.e.\ in the~${\bf{15}}$. Hence we have additionally included $S_{15}^{}$ and
$T^{(\prime)}_{15}$ in Table~\ref{tab:fullfieldcontent}, where the
transformation properties are identical to $S$ and $T^{(\prime)}$ of the
original model with the exception of $SU(4)$. Since the original flavon fields
were taken to be $SU(4)$ singlets, we have written $S_1=S$ and
$T^{(\prime)}_1=T^{(\prime)}$. The resulting changes to the Yukawa
interactions are given by simple replacements in
Eq.~\eqref{eq:fermionlagrangian} such as for instance\footnote{Note that only
  $S_1^\dagger$ can couple in the term of the third line of
  Eq.~\eqref{eq:fermionlagrangian-nu}.} 
\be
\ol \Sigma_L \, (\kappa_S S) \, \Sigma_R
~~ \rightarrow ~~
\ol \Sigma_L \, (\kappa_{S_1} S_1+ \kappa_{S_{15}} S_{15}) \, \Sigma_R \ .
\ee
Inserting the flavon VEVs and dividing by the new-physics (NP) mass scale $M$,
we can define 
dimensionless $3\times 3$ matrices $s_i=\kappa_{S_i} \langle S_i \rangle /M$. 
The presence of the adjoint entails a difference between quarks and leptons:
the $SU(3)$ conserving direction of the ${\bf{15}}$, written as a matrix, is
Diag\,$(1,1,1,-3)$ so that the leptons pick up a relative factor of $-3$
compared to the quarks.\footnote{This so-called Georgi-Jarlskog factor
was first discussed in the context of $SU(5)$ in~\cite{Georgi:1979df}, where the
scalar sector was enlarged by a Higgs multiplet in the~${\bf{45}}$ which
(contrary to the standard Higgs in the~${\bf{5}}$) treats quarks and leptons
differently.}
Hence we are led to the following replacements 
\be
\begin{array}{lll}
\text{quarks:} & s~~\rightarrow~~ s_q~ \equiv~ s_1+s_{15} \,, ~ \quad
& t'~~\rightarrow~~ t'_q~ \equiv~ t'_1+t'_{15} \, , \\[3mm]
\text{leptons:} ~~~ & s~~\rightarrow~~ s_l~ \equiv ~s_1-3\,s_{15} \,, ~ \quad
& t'~~\rightarrow~~ t'_l~ \equiv ~t'_1-3\,t'_{15} \, .~~~~~
\end{array}
\ee
A setup where all $SU(4)$ singlet flavon fields are accompanied by 
flavons in the adjoint of $SU(4)$ therefore decouples the flavour structure of
the quark and lepton sector completely. Having discussed the phenomenology of
the quark sector extensively in~\cite{Feldmann:2015zwa}, we focus exclusively
on the leptons in the following. However, we do not pursue a numerical fit to
neutrino oscillation data because the implementation of light Majorana
neutrino masses in our model introduces the independent flavour structure of
$\langle S_\nu^{}\rangle$ so that a realistic PMNS mixing matrix can always
be achieved regardless of quark and charged-lepton flavour.




Analogously to the quark sector, bilinear mass terms of the charged leptons,
including their heavy partners, can be expressed in terms of a $9\times 9$
matrix. Defining the basis\footnote{We use the label $\ell$ for charged
leptons while $l$ refers to both charged and neutral leptons.}
\be \label{eq:defPSI}
{\overline \Psi^\ell_L} ~\equiv ~(\overline q_L^\ell , \overline \Sigma_L^\ell , 
\overline \Xi_L^\ell) \ , ~ \qquad
{\Psi^\ell_R} ~\equiv ~( q_R^\ell ,  \Sigma_R^\ell ,  \Xi_R^\ell) \ ,
\ee
we have
\be\label{eq:Me}
\mathcal M^\ell ~=~ 
\begin{pmatrix} 
0 & \id \,  \lambda \epsilon_d & \id \\
\id  & s_l- t'_l  & 0  \\
\id\,  \lambda  \epsilon_d  & 0 & s_l
\end{pmatrix} M \ ,
\ee 
where 
\be
\epsilon_d ~\equiv ~ \frac{v_d }{\sqrt{2}\,M} 
~=~ \frac{v}{\sqrt{2}\,\sqrt{1+\tan^2\beta}\,M}\ ,
\ee
and $v=246$~GeV. In the limit where $s_l^{}, t'_l \gg 1$, the three light
charged-lepton masses can be obtained by integrating out the heavy degrees of
freedom. A straightforward calculation gives rise to an effective Yukawa
matrix $Y_\ell$ of the form
\be
\label{eq:Yell-eff}
Y_\ell ~\simeq -\lambda \left[   \frac{1}{s_l- t'_l} \,+\, \frac{1}{s_l}
  \right] \ .
\ee
In order to assess the validity of this approximate formula,
let us consider the one-generation case. Assuming $s_l \sim t'_l$,
the determinant of $\m M^\ell$ is of order $M^2 s_l v_d$. On the other hand, 
the product of the singular values reads $m_\tau M^2 s_l^2$, where $m_\tau$
is the tau lepton mass and the heavy  partners are assumed to  have a mass of
order $M s_l$.  Comparing these two expressions for $|\mathrm{Det}\,\mathcal
M^\ell|$ provides an estimate of the magnitude of $s_l \sim v_d/m_\tau$ which
is greater than about 10 for $\tan\beta\in [1,15]$.  Hence, $s_l^{}, t'_l \gg
1$ is generally satisfied so that Eq.~\eqref{eq:Yell-eff} provides a good
approximation, even for the third generation.

The flavour structure of the charged leptons originates in the matrices $s_l$
and $t'_l$. These are free parameters of the theory which, however, have to be
chosen such as to yield correct charged-lepton masses. It is therefore convenient
to  replace $s_l- t'_l$ in Eq.~\eqref{eq:Me} in favour of $Y_\ell$ using
Eq.~\eqref{eq:Yell-eff}. Going to a basis in which $Y_\ell$ is 
diagonal we have
\be
s_l- t'_l ~\rightarrow~ - \left[\frac{\hat Y_\ell}{\lambda} + s^{-1}_l
  \right]^{-1}~=~
- \left[\frac{\sqrt{2}\,\hat M^\ell}{\lambda\,v_d} + s^{-1}_l
  \right]^{-1}
\ ,\label{eq:replacement}
\ee
where hats denote diagonal $3\times 3$ matrices and $\hat M^\ell =
\text{Diag}\,(m_e,m_\mu,m_\tau)$ contains the measured charged-lepton masses.
Having eliminated $t'_l$ in Eq.~\eqref{eq:Me}, the charged-lepton masses are
automatically correctly described, regardless of the flavour structure encoded in $s_l$. 
The latter provides the dominant source for lepton-flavour violation in our model.




\section{Charged-lepton flavour sector\label{sec:3}}
\cleqn

In order to discuss the flavour phenomenology of the charged leptons, we have
to express the gauge-kinetic terms in the mass basis. The procedure is
analogous to the treatment of the quark sector. However there exist some
simplifications due to the smallness of the tau lepton Yukawa coupling, 
as well as subtle differences resulting from the fact that
neutrinos can be regarded as massless particles for our purposes. 


\subsection{\label{sec:diag-ell}Diagonalising  the charged-lepton mass matrix}

Our starting point is the $9\times 9$ mass matrix of Eq.~\eqref{eq:Me} which
we rewrite using the singular value decomposition
\be\label{eq:para-ell}
s_l= V_s^\dagger \, \hat s \, U_s \ , \qquad
s_l-t'_l ~=~ V_{t}^\dagger \, \hat t  \, U_{t}  \ .
\ee
Here $V_{s,t}$ and $U_{s,t}$ are unitary matrices, while $\hat s$ and $\hat t$
denote the diagonalised versions of $s_l$ and $s_l-t'_l$, respectively. The
latter is related to the charged-lepton masses via
Eq.~\eqref{eq:replacement}.  Hence, $V_s$, $U_s$ and $\hat s$ can be regarded
as free parameters, while $V_t$, $U_t$ and $\hat t$ are derived from the bi-unitary
diagonalisation of the right-hand side of Eq.~\eqref{eq:replacement}.  

In the following we briefly sketch the sequence of basis transformations
which diagonalises $\m M^\ell$ of Eq.~\eqref{eq:Me}. More details are provided
in Appendix~\ref{app:diagonalize} where we follow closely the corresponding
discussion in~\cite{Feldmann:2015zwa}. First, we 
absorb the $3\times 3$ matrices $V_{s,t}$ and $U_{s,t}$ into a redefinition of
the fields such that the only non-diagonal blocks within $\mathcal M^\ell$ are
the ones proportional to $\epsilon_d$. As shown in
Appendix~\ref{app:diagonalize}, the resulting matrix can be easily
diagonalised in the limit of $\epsilon_d =0$. Applying this basis 
transformations on the full mass matrix $\mathcal M^\ell$ yields
\bea \label{eq:Mell-basis3}
\mathcal M^\ell &\rightarrow& 
\begin{pmatrix} 
a\,\epsilon_d & b\,\epsilon_d & 0 \\
0 & \hat e & 0 \\
c\,\epsilon_d & d\,\epsilon_d & \hat f
\end{pmatrix} M \ ,
\eea
where $a$, $b$, $c$, $d$, $\hat e$ and $\hat f$ are $3\times 3$ matrices whose
definition can be found in Appendix~\ref{app:diagonalize}. A further, more complicated
transformation which block-diagonalises $\mathcal M^\ell$ to second order in
$\epsilon_d$ is given explicitly in Appendix~\ref{app:diagonalize}. It results
in the following simple form of $\mathcal M^\ell$,
\bea \label{eq:Mell-basis4}
\mathcal M^\ell &\rightarrow& 
\begin{pmatrix} 
a\,\epsilon_d & 0 & 0 \\
0 & \hat e + \m O(\epsilon_d^2)& 0 \\
0 & 0 & \hat f + \m O(\epsilon_d^2)
\end{pmatrix} M  \:+~ \m O(\epsilon_d^3) \, M\ .
\eea
The final step of the sequence of basis transformations diagonalises the upper
left block $a v_d /\sqrt{2}$  of $\m M^\ell$, which gives rise to the diagonal
charged-lepton Yukawa coupling 
\be
\hat Y_\ell ~=~ \m V ~a ~\m U^\dagger \ . 
\ee
The unitary $3\times 3$ matrices $\m V$ and $\m U$ are directly related to the
parametrisation in Eq.~\eqref{eq:para-ell}. This becomes clear by considering
the explicit form of $a$, which is calculated in Appendix~\ref{app:diagonalize}
to second order in $\hat s^{-1}$ and $\hat t^{-1}$. From Eq.~\eqref{eq:a},
together with Eq.~\eqref{eq:para-ell}, we get
\bea
a &\approx & 
- \lambda \, U^{}_s\,(U^\dagger_s\;\hat s^{-1}\, V_s^{}  + U^\dagger_t\; \hat
t^{-1}\, V_t^{} ) \, V_t^\dagger
~=~
- \lambda \, U^{}_s\,\left(\frac{1}{s_l} + \frac{1}{s_l-t'_l} \right) \,
V_t^\dagger \ .
\eea
Inserting the substitution of Eq.~\eqref{eq:replacement} shows that 
\be
\m V ~\approx~ U^\dagger_s \ , \qquad
\m U ~\approx~ V^\dagger_t \ .
\ee
These relations represent a significant simplification compared to the quark
sector where $\m V$ and $\m U$ could only be determined numerically. 
All steps of the sequence of basis transformations are explicitly stated in
Appendix~\ref{app:diagonalize}. They describe the change from the original
flavour basis $\Psi^\ell_{L,R}$ to the approximate mass basis
$\Psi'^\ell_{L,R}$. Before applying these transformations to the gauge-kinetic
terms, it is necessary to consider the relevant transformations in the neutral sector.


\subsection{The neutral-lepton mass matrix}

In order to study charged-lepton flavour violation, it is sufficient to treat
the left-handed neutrinos in the massless limit. The advantage of this limit lies in the
possibility to choose identical basis transformations for both components of
the lepton doublet. However, due to the mixing with the heavy partners
of the neutrinos, the massless neutral fermions do not simply correspond to
$\ol q^\nu_L \subset \ol \Psi^\nu_L$. It is therefore necessary to scrutinise
the mass matrix of the neutral fermions in more detail. In addition to the 18
neutral components of $\ol \Psi^\nu_L$ and $\Psi^\nu_R$, the model presented
in~\cite{Feldmann:2015zwa} introduces two further fermions $\ol\Theta_L$ and
$\Theta_R$ which acquire very large Majorana masses at the  flavour symmetry
breaking scale $\Lambda_\nu^{(\prime)}$,
cf. Table~\ref{tab:fullfieldcontent}. The coupling of $\ol\Theta_L$ to
$\Sigma_R$ induces a Majorana mass around the seesaw scale for the latter. Hence,
integrating out $\ol\Theta_L$ and $\Theta_R$, we obtain a bilinear mass term
of the form
\be
 \frac{1}{2} ~  {\Psi_{\mathrm{Maj}}^T} ~ \mathcal M^{\mathrm{Maj}} ~\Psi_{\mathrm{Maj}} \ ,
\ee
where 
\be\label{eq:Maj-vector}
\Psi_{\mathrm{Maj}} ~=~
({\ol\Psi_L^\nu}^T , \Psi_R^\nu)
~=~ ({\ol q_L^\nu}^T,{\ol \Sigma_L^\nu}^T,{\ol \Xi_L^\nu}^T, q_R^\nu ,
\Sigma_R^\nu , \Xi_R^\nu) \ ,
\ee
and
\be
\mathcal M^{\mathrm{Maj}} ~=~ \begin{pmatrix} 
0&0&0&0& \id  \lambda\epsilon_u & \id \\
0&0&0&\id& s_l+t'_l & 0 \\
0&0&0&\id \lambda \epsilon_u & 0 & s_l \\
0&\id & \id \lambda \epsilon_u &0&0&0\\
\id \lambda\epsilon_u & (s_l+t'_l)^T & 0&0&y&0\\
\id&0&s_l^T&0&0&0
\end{pmatrix}M\ .\label{eq:18x18}
\ee
The only direct Majorana-type entry of the  $18\times 18$  mass matrix
$\mathcal M^{\mathrm{Maj}}$ is given by, cf. Appendix~A.1.1 of~\cite{Feldmann:2015zwa},
\be y \equiv
\frac{M_{\Sigma^\nu_R}}{M} = \frac{(\varphi'_\alpha \Lambda_\varphi)^2}{M
  \Lambda_\nu} s_\nu^{-1} \ .
\ee
In the following we intend to identify the directions of the three lightest
neutral fermions. These correspond to the massless states of the limit
$\epsilon_u= 0$. Having diagonalised the submatrix $s_l$ of 
$\mathcal M^{\mathrm{Maj}}$ analogously to the charged sector by absorbing
$V_s$ and $U_s$ into a redefinition of the fields $q^\nu_L$ and
$\Xi^\nu_{L,R}$, it is straightforward to rotate the $\ol q^\nu_L \,
\Xi^\nu_R$ coupling into the $\ol\Xi^\nu_L\, \Xi^\nu_R$ mass term. As detailed
in Appendix~\ref{app:majorana}, such a basis transformation simplifies the
mass matrix of Eq.~\eqref{eq:18x18}, with $\epsilon_u=0$, to
\be
\mathcal M^{\mathrm{Maj}}\Big|_{\epsilon_u= 0} ~\rightarrow~ \begin{pmatrix} 
0&0&0&0& 0 & 0 \\
0&0&0&\id& s_l+t'_l & 0 \\
0&0&0&0 & 0 & \hat s \\
0&\id & 0 &0&0&0\\
0 & (s_l+t'_l)^T & 0&0&y&0\\
0&0&\hat s&0&0&0
\end{pmatrix}M\ .\label{eq:18x18simpl}
\ee
In this basis, the first three components correspond to the massless neutrinos
which decouple from the massive neutral fermions. Furthermore, we find 
separated Dirac pairs whose masses are given by the diagonal matrix $\hat s$;
their entries are all larger than $M$. The remaining degrees of freedom mix via 
\be
\begin{pmatrix} 
0&\id& s_l+t'_l \\
\id & 0&0\\
 (s_l+t'_l)^T &0&y
\end{pmatrix}M\ ,\label{eq:9x9simpl}
\ee 
resulting in three masses around the seesaw scale $yM=M_{\Sigma^\nu_R}$ as
well as six masses of order~$M$.
Having isolated the three massless neutrinos from the massive neutral
fermions in Eq.~\eqref{eq:18x18simpl}, no further transformation which mixes
light and heavy degrees of freedom must be applied on $\Psi_{\mathrm{Maj}}$.
The only allowed additional transformations are mixings of the three
neutrinos themselves. Being massless, it is convenient to choose these
identical to the corresponding unitary transformations of the charged leptons.


\subsection{Gauge-kinetic terms}

The basis transformations discussed above must now be applied to the
gauge-kinetic terms. As the flavour gauge bosons are far too heavy to have an
impact on experimental observables, we do not consider them in the
following. On the other hand, among the PS gauge bosons, only the SM ones are
relevant for low-energy phenomenology. A comprehensive discussion of the 
gauge-kinetic terms involving the fermionic PS multiplets can be found
in~\cite{Feldmann:2015zwa}. In the original basis, it applies to the lepton
sector without any modifications. However, the change from the flavour to the
mass basis differs for quarks and leptons. We therefore have to reanalyse the
flavour structure of the neutral and charged currents for leptons.  

The gauge-kinetic terms involving the electroweak gauge bosons have
been derived in~\cite{Feldmann:2015zwa}. Although we are not interested in neutral
currents involving the neutrinos, we include these in the following in order to 
facilitate a direct comparison with the expressions of the quark sector. In
the original flavour basis, we have\footnote{We use the standard abbreviation
$s_W=\sin \theta_W$ and $c_W=\cos\theta_W$ for the sine and cosine of the 
weak mixing angle.}
\begin{align}
\mathcal{L}_\text{kin} ~ \supset 
& ~~~~\: \frac{g}{c_W}\;\,{\overline \Psi^{}_L} \biggl( \left(\tau^3\, - s_W^2 Q_e\right)\id - {\mathcal K'_L}\,\tau^3 \biggr)\slashed{Z} {\Psi^{}_L} \nonumber\\
&+\
\frac{g}{c_W}\;\,{\overline \Psi^{}_R} \bigg( ~~-\,s_W^2\;Q_e \id \hspace{4.5mm}
+ {\mathcal K_R}  \, \tau^3\bigg)\slashed{Z}{\Psi^{}_R} \nonumber\\
& +\ 
\frac{g}{\sqrt{2}} \;\, {\overline \Psi^{\nu}_L} \,{\mathcal K_L}\,
\slashed{W}^+ {\Psi^{\ell}_L}  + \mathrm{h.c.}\nonumber\\
&+\ 
\frac{g}{\sqrt{2}} \;\, {\overline \Psi^{\nu}_R}\, {\mathcal K_R}\,
\slashed{W}^+ {\Psi^{\ell}_R}+ \mathrm{h.c.}
 \,,\label{eq:GRVbrokenkinlag}
\end{align}
where 
\be
{\mathcal K'_L} = 
\begin{pmatrix} 0&0&0 \\ 0&\id &0 \\0&0&0 \end{pmatrix} \ , 
\qquad
{\mathcal K_R} = 
\begin{pmatrix} 0&0&0 \\ 0&0&0 \\0&0&\id \end{pmatrix} \ ,
\qquad
{\mathcal K_L} = 
\begin{pmatrix} \id&0&0 \\ 0&0 &0 \\0&0&\id \end{pmatrix} \ .
\ee
Going to the mass basis, only the terms proportional to $\mathcal K'_L$,
$\mathcal K_R$ and $\m K_L$ change their form. The explicit results are
given in Appendix~\ref{app:gaugekin}. Focusing on the upper left $3\times 3$
blocks, we obtain
\begin{align}\label{eq:L_kingauge}
\mathcal{L}_\text{kin} ~ \supset 
&\phantom{\:\:+~}
\frac{g}{c_W}\, {\overline {\ell}_L}\biggl( \left(-\mbox{$\frac{1}{2}$}\, - s_W^2
Q_e\right)\id +\mbox{$\frac{1}{2}$} \, \Delta g_{Z\overline{\ell}^{}_L\ell^{}_L}\biggr)\slashed{Z}\, {{\ell}_L} \nonumber\\
& +~
\frac{g}{c_W}\, {\overline {\ell}_R} \biggl(~~ -\,s_W^2\:Q_e
\id \hspace{6.3mm}-\mbox{$\frac{1}{2}$}\,\Delta
g_{Z\overline{\ell}^{}_R\ell^{}_R} 
\biggr)\slashed{Z}\, {{\ell}_R} \nonumber\\
& +~
 \frac{g}{\sqrt{2}}\;{\overline {\nu}_L} \left(\id - \Delta g^{}_{W\ol\nu_L\ell_L} \right)
 \slashed{W}^+{{\ell}_L} + \mathrm{h.c.} \ ,
\end{align}
where the mass basis $q'^{\ell}_{L,R}$ and $q'^\nu_L$ has been renamed by 
$\ell_{L,R}$ and $\nu_L$, respectively. The anomalous $Z$ and $W$ couplings
are given as
\bea
\Delta g_{Z\ol\ell_L\ell_L}&=&
U_t^\dagger \hat t^{-2} U_t \, \lambda^2 \,\epsilon_d^2 =
\left[\frac{\sqrt{2} \hat M^\ell}{\lambda\,v_d}+s_l^{-1}\right]  
\left[\frac{\sqrt{2} \hat M^\ell}{\lambda\,v_d}+s_l^{-1}\right]^\dagger \, \lambda^2
\,\epsilon_d^2 \   ,~~~~~
\label{eq:gZlLlL}
\\
\Delta g_{Z\ol\ell_R\ell_R}&=&
V_s^\dagger \hat s^{-2} V_s \, \lambda^2 \,\epsilon_d^2 =
\big[s_l^{-1}\big]^\dagger \big[s_l^{-1}\big] \, \lambda^2
\,\epsilon_d^2  \ ,
\label{eq:gZlRlR}
\\
\Delta g_{W \ol\nu_L \ell_L} & =&
\frac{1}{2} \, U_t^\dagger \hat t^{-2} U_t \, \lambda^2 \,\epsilon_d^2 
=\frac{1}{2} \,
\left[\frac{\sqrt{2} \hat M^\ell}{\lambda\,v_d}+s_l^{-1}\right]  
\left[\frac{\sqrt{2} \hat M^\ell}{\lambda\,v_d}+s_l^{-1}\right]^\dagger \, \lambda^2
\,\epsilon_d^2 \  . ~~~~~
\label{eq:Wnul}
\eea
A few comments are in order. In Eq.~\eqref{eq:L_kingauge}, we have omitted the
neutral currents of the neutrinos. Furthermore, we do not show the charged
currents involving the right-handed neutrinos as $q'^\nu_R$ corresponds to 
heavy neutral fermions. Working in the limit of massless left-handed neutrinos,
we do not encounter a non-trivial PMNS matrix in the corresponding charged current.
Finally, note that the anomalous $W$ coupling is related to the anomalous $Z$
coupling by $\Delta g_{W \ol\nu_L \ell_L} =\frac{1}{2} \Delta g_{Z \ol\ell_L \ell_L}$.


\subsection{Anomalous Higgs coupling}

The discussion of the anomalous coupling of the charged leptons to the Higgs
proceeds analogously to the treatment of the Higgs-quark-quark coupling in
Section~3.3 of~\cite{Feldmann:2015zwa}. It requires the basis transformation
to be computed to third order in the expansion
parameter~$\epsilon_d$. Building on the results of~\cite{Feldmann:2015zwa}, we 
parameterise the relevant higher-order correction by
\bea
\Delta g_{h\overline{\ell}\ell} &\approx&  -~ \epsilon_d^2 \, 
\Bigl[\,
  U_s^\dagger \,  b\,\hat{e}^{-2}_{}\,b^\dagger\, U_s \cdot \hat Y_\ell 
+  \hat Y_\ell \cdot\, V_t^\dagger   \,c^\dagger\,\hat f^{-2}_{}\,c \,  V_t 
\,\Bigr]\label{eq:anomal-Higgs}
\\
&\approx&  -~ \lambda^2\,\epsilon_d^2 \,
\Bigg\{
\left[\frac{\sqrt{2} \hat M^\ell}{\lambda\,v_d}+s_l^{-1}\right]  
\left[\frac{\sqrt{2} \hat M^\ell}{\lambda\,v_d}+s_l^{-1}\right]^\dagger \!\!
 \cdot \hat Y_\ell 
+  \hat Y_\ell \cdot\, \big[s_l^{-1}\big]^\dagger \big[s_l^{-1}\big] 
\,\Bigg\} .~~~~\nonumber
\eea
Performing the sequence of basis transformations discussed in
Section~\ref{sec:diag-ell}, the Yukawa matrix is given by~\cite{Feldmann:2015zwa}  
\be
  \label{eq:GRVYukeps2}
  Y_\ell ~\approx ~ \hat Y_\ell ~+~ \frac{1}{2} \,  \Delta g_{h\overline{\ell}\ell} 
\, ,
\ee
while the Higgs-lepton-lepton coupling takes the form
\be
  g_{h\overline{\ell}\ell} ~\approx ~  
 \hat Y_\ell ~+~ \frac{3}{2} \,  \Delta g_{h\overline{\ell}\ell} 
\, .
\ee
The latter is thus related to the former via
\bea
 g_{h\overline{\ell}\ell} ~\approx ~  Y_\ell ~+~ \Delta g_{h\overline{\ell}\ell} 
\,.\label{eq:an-Higgs}
\eea
The diagonalisation of $Y_\ell$ in Eq.~\eqref{eq:GRVYukeps2} to this higher
order in $\epsilon_d$ is achieved by unitary matrices which deviate from the
identity by contributions of order $\epsilon_d^2$. Such a basis change does
not affect the second term in Eq.~\eqref{eq:an-Higgs} at the given
order. Hence, the deviation of the Higgs coupling from the diagonal
charged-lepton Yukawa matrix is simply given by $\Delta
g_{h\overline{\ell}\ell}$ of Eq.~\eqref{eq:anomal-Higgs}. 




\section{LFV observables from  effective field theory\label{sec:4}}
\cleqn

In this section we discuss the effects of our model on 
the low-energy LFV observables. In particular, these are 
induced by flavour-changing couplings of the fermions 
to the SM gauge bosons~($Z,W$) and Higgs particle~($h$), while the 
couplings to the new gauge bosons ($Z'$ etc.) are 
additionally suppressed due to their heavy mass, as 
we have already explained in~\cite{Feldmann:2015zwa}.
Moreover, the mass scale associated with 
the dimension-five Weinberg operator, relevant for the 
neutrino masses, is much larger than the NP scale $M$ 
for charged LFV in our model, and the respective heavy Majorana neutrinos 
decouple. The relevant operators 
in an effective theory at the electroweak scale can then be identified as
\begin{align}
 \mathcal{L}_{\mathrm{LFV}}\, = ~&\,\frac{g  Z^\mu}{2\,c_W}\left(  \Delta g^{ij}_{Z \ol \ell _L \ell_L}    (\ol \ell_i \gamma_\mu P_L \ell_j)
 - \Delta g^{ij}_{Z \ol \ell _R \ell_R} (\ol \ell_i \gamma_\mu P_R \ell_j) \right) \nn \\
 &-\,
\frac{g W^\mu}{\sqrt{2}} \, \Delta g^{ij}_{W \ol \nu _L \ell_L}  (\ol
 \nu_i \gamma_\mu P_L \ell_j)
\,+\, \frac{h}{\sqrt{2}} \, \Delta g^{ij}_{h \ol \ell \ell}  (\ol \ell_i
 P_R \ell_j)  \,+\,  \mathrm{h.c.} \,,
\label{eq:LFV} 
\end{align}
where the $3\times 3$ coupling matrices $\Delta g_{Z \ol \ell _L \ell_L}$, $\Delta g_{Z
  \ol \ell _R  \ell_R}$, $ \Delta g_{W \ol \nu _L \ell_L}$ and $\Delta g_{h \ol \ell \ell}$
are given in Eqs.~(\ref{eq:gZlLlL}--\ref{eq:anomal-Higgs}).
The terms in Eq.~\eqref{eq:LFV} descend from the gauge-invariant 
dimension-six operators 
$$
\Phi^\dagger i \overleftrightarrow{D}^\mu \Phi \, (\ol E_i \gamma_\mu
E_j) \,, \quad 
\Phi^\dagger i \overleftrightarrow{D}^\mu \Phi \, (\ol L_i \gamma_\mu
L_j) \,, \quad 
\Phi^\dagger i \overleftrightarrow{ \tau^A D}^\mu \Phi  \, (\ol L_i
\tau^A \gamma_\mu L_j)\,, \quad 
(\Phi^\dagger\Phi) \, \ol L_i \Phi  E_j \,,$$
appearing in the Buchm\"uller-Wyler 
Lagrangian~\cite{Grzadkowski:2010es, Buchmuller:1985jz}
after electroweak symmetry breaking.

In the following, we will focus on radiative transitions of the type 
$\ell_i \to \ell_f \gamma $, tri-lepton decays $\ell_i \to 3 \ell_f$ and 
lepton conversion in nuclei. 
As the energy release of each process is typically of the order of 
the mass of the initial charged lepton, we consider -- as usual --
a low-energy effective Lagrangian where all fields with masses 
above the charged-lepton mass have been integrated out,
notably the heavy SM gauge bosons.


\subsection{The decay $\boldsymbol{\mu \to e \gamma}$}

\begin{figure}[t!pb]
 \begin{center}
\includegraphics[width=1.0\textwidth]{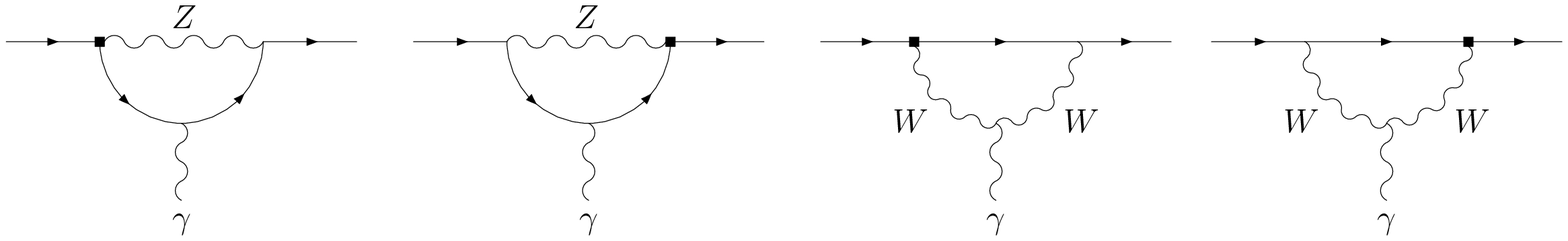} 
 \end{center}
\caption{1-loop topologies contributing to $A_{L/R}$. The black square
indicates the insertion of an anomalous $Z$ or $W$ coupling.}
 \label{fig_ARL}
\end{figure}

Starting with the decay $\mu \to e \gamma$,
we follow the conventions of 
\cite{Beneke:2015lba} and consider the low-energy operators
\begin{eqnarray}
\label{LowEngergyMuEgamma}
\mathcal{L}_{\ \mu\to e \gamma } &=&
\,
A_R \, m_\mu \, F_{\sigma\rho} \, (\ol \ell_e \sigma^{\sigma\rho} P_R \ell_\mu)  + 
A_L \, m_\mu \, F_{\sigma\rho} \, (\ol \ell_e \sigma^{\sigma\rho} P_L
\ell_\mu)  +\,\text{h.c.} \ .
\end{eqnarray}
With this definition the branching ratio of $\mu \to e \gamma$ can be written as~\cite{Beneke:2015lba}
\begin{equation}
\label{eq:BrMEG}
{\rm{Br}}(\mu \to e\gamma)=  
\frac{m_\mu^5}{4\pi \, \Gamma_\mu}(|A_L|^2+|A_R|^2)\,. 
\end{equation}
The coefficients $A_{L/R}$ receive contributions from 
1-loop diagrams involving anomalous $Z$ and $W$ couplings, see Fig.~\ref{fig_ARL}.
For completeness, we have also included 2-loop diagrams of the ``Barr-Zee
type''~\cite{Barr:1990vd,Chang:1993kw} which involve the anomalous Higgs
couplings as these can be dominant in some corners of parameter space.
Using the results of~\cite{Beneke:2015lba}, the coefficients $A_{L/R}$ 
can be expressed in terms of the various anomalous couplings of 
Eq.~\eqref{eq:LFV}, and we obtain
\begin{align}\label{ARtotal}
m_\mu A_R = 
&\;-\frac{2 Q_\ell e}{3(4 \pi)^2 v^2}
\Bigg( 
s_W^2 \left[
-m_\mu \Delta g^{12}_{Z \ol \ell _L \ell_L} 
+m_e \Delta g^{12}_{Z \ol \ell _R \ell_R}
\right] 
 - m_\mu \Delta g^{12}_{Z \ol \ell _L \ell_L}   
-\frac{3}{2} m_e \Delta g^{12}_{Z \ol \ell _R \ell_R}\nn\\
& \hspace{27mm}-\frac{5}{2}\, m_\mu \Delta g^{12}_{W\ol\nu_L\ell_L} \Bigg)
   ~+~ A_{BZ}
\,\frac{1}{\sqrt{2}}\,
\Delta g^{12}_{h \ol \ell \ell} \ ,
\\
\label{ALtotal}
 m_\mu A_L =  
 &\;-\frac{2Q_\ell e}{3(4 \pi)^2 v^2}
\Bigg( 
s_W^2  \left[
-m_e \Delta g^{12}_{Z \ol \ell _L \ell_L}
+m_\mu \Delta g^{12}_{Z \ol \ell _R \ell_R}
\right] 
- m_e  \Delta g^{12}_{Z \ol \ell _L \ell_L}   
-\frac{3}{2} m_\mu \Delta g^{12}_{Z \ol \ell _R \ell_R}
\nn \\
&\hspace{27mm} 
-\frac{5}{2}\, m_e \Delta g^{12}_{W\ol\nu_L\ell_L} \Bigg)
 ~ +~ A_{BZ}
\,\frac{1}{\sqrt{2}}\,
[\Delta g^{\dagger}_{h \ol \ell \ell}]^{12}\ ,
\end{align}
where the expression for the Barr-Zee coefficient $A_{BZ}$ 
can also be found in~\cite{Beneke:2015lba}.
The adaptation to other transitions ($\tau \to \mu\gamma$, 
$\tau \to e\gamma$) is straightforward. 
Note that, in practice, all terms
proportional to the electron mass $m_e$ can be dropped in
Eqs.~(\ref{ARtotal},\ref{ALtotal}). 

\subsection{The decay $\boldsymbol{\mu \to 3 e}$}

Turning to the decay $\mu \to 3 e$,
we have to add to Eq.~(\ref{LowEngergyMuEgamma}) a set of
four-lepton operators (following again the notation of~\cite{Beneke:2015lba}),
\begin{eqnarray}
\label{LowEngergyMu3e}
\mathcal{L}_{\ \mu\to 3e} &=&
\mathcal{L}_{\ \mu\to e \gamma} \nn \\
&&+ \,g_1 \,(\ol \ell_e P_R \ell_\mu) \; (\ol \ell_e  P_R \ell_e) 
+ g_2 \, (\ol \ell_e  P_L \ell_\mu)\; (\ol\ell_e P_L \ell_e)  
\nn \\ 
&&+\,g_3 \, (\ol \ell_e \gamma^\nu P_R \ell_\mu)\;(\ol \ell_e 
\gamma_\nu P_R \ell_e) 
+  g_4 \,(\ol \ell_e \gamma^\nu P_L\ell_\mu) \; (\ol \ell_e \gamma_\nu 
P_L \ell_e) 
\nn \\
&&   +\,g_5 \, (\ol \ell_e \gamma^\nu P_R \ell_\mu)\;(\ol \ell_e 
\gamma_\nu P_L \ell_e)
+  g_6 \, (\ol \ell_e \gamma^\nu P_L \ell_\mu)\;(\ol \ell_e 
\gamma_\nu P_R \ell_e) +\,\text{h.c.} \ .
\end{eqnarray}
The branching ratio for 
$\mu \to 3e$ can then be expressed as 
\begin{eqnarray}
\label{eq:Brmu3e}
{\rm Br}(\mu \to 3 e) &=& \frac{m_\mu^5}{1536 \,\pi^3 \, \Gamma_\mu}
\bigg[\frac{|g_1|^2+|g_2|^2}{8}+2 \left(|g_3|^2+|g_4|^2\right)+|g_5|^2+|g_6|^2 
\nn \\[0.1cm]
&&  \phantom{\frac{m_\mu^5}{1536 \,\pi^3 \, \Gamma_\mu}}
- \,8 e\, {\rm{Re}}\left[A_R \, (2 g_4^*+g_6^*)+A_L\, (2 g_3^*+g_5^*) \right]
\nn \\[0.1cm]
&&\phantom{\frac{m_\mu^5}{1536 \,\pi^3 \, \Gamma_\mu}}
+\,64 e^2 \left(\ln\frac{m_\mu}{m_e}-\frac{11}{8}\right)(|A_L|^2+|A_R|^2) \bigg].
\end{eqnarray}
Again, the adaptation to different flavour transitions is straightforward. 
Notice that the coefficients $A_{L/R}$ only arise at 1-loop level, while the
couplings $g_{i}$ are generated by tree-level exchange of weak gauge
bosons.\footnote{Our Pati-Salam model also generates contributions to the
coefficients $g_{1,2}$ via the anomalous Higgs coupling, which is similar to
the situation in the Randall-Sundrum scenario discussed in~\cite{Beneke:2015lba}. 
However, as noted in~\cite{Beneke:2015lba}, the associated tree-level matching
diagrams are suppressed by a factor of the electron mass and can therefore be
neglected.} 
However, as can be seen from Eq.~\eqref{eq:Brmu3e}, $A_{L/R}$ come 
with a large pre-factor and should thus be included for completeness. 
Expressing the couplings $g_i$ of the four-lepton operators in terms of the
anomalous $Z$ and $W$ couplings to fermions of Eq.~\eqref{eq:LFV}, we find 
\begin{align}
\label{eq:gCoeffs}
g_1&\simeq g_2 \simeq 0 \ ,\cr 
g_3&=   \frac{2s_W^2}{v^2}  \Delta g^{12}_{Z \ol \ell _R \ell_R}  \ ,   \qquad 
\hspace{7.2mm}
g_4=   -\frac{2 s_W^2-1}{v^2} \Delta g^{12}_{Z \ol \ell _L \ell_L} \ ,\cr 
g_5&=   \frac{2 s_W^2-1}{v^2} \Delta g^{12}_{Z \ol \ell _R \ell_R} \ , \qquad 
g_6=  - \frac{2s_W^2}{v^2} \Delta g^{12}_{Z \ol \ell _L \ell_L}  \ .
\end{align}


\subsection{Muon conversion in nuclei}

Finally, the appropriate low-energy Lagrangian for muon conversion in nuclei now also 
contains mixed four-fermion operators with lepton
and quark currents (see e.g.~\cite{Kuno:1999jp,Crivellin:2013ipa}),
\begin{eqnarray}
\label{LowEngergyMuonconversion}
\mathcal{L}_{\mu N\to e N} &\!=\!& \mathcal{L}_{\mu \to e\gamma} \nn\\
&&  + \,\sum_{q=u,d,s} \frac{m_q m_\mu}{M_h^2} \,  c^q_{SL} \,   
(\ol \ell_e   P_R \ell_\mu)( \ol q q)
+ \sum_{q=u,d,s} \frac{m_q m_\mu}{M_h^2} \, c^q_{SR} \,( \ol \ell_e   
P_L \ell_\mu)( \ol q q )
\nn  \\         
&& + \,
\sum_{q=u,d} c^q_{VL}  \,( \ol \ell_e  \gamma^\nu P_L \ell_\mu)( \ol q  
\gamma_\nu q) 
+
\sum_{q=u,d} c^q_{VR}  \, (\ol \ell_e  
\gamma^\nu P_R \ell_\mu)( \ol q  \gamma_\nu q)
\nn \\
&&  + \,\frac{\alpha_s m_\mu}{M_h^2} \, c^L_{gg} \, (\ol \ell_e   
P_R \ell_\mu) \, G^{A,\sigma\rho}G^A_{\sigma\rho}
+ \frac{\alpha_s m_\mu}{M_h^2} \, c_{gg}^R \, (\ol \ell_e   
P_L \ell_\mu)\, G^{A,\sigma\rho}G^A_{\sigma\rho}
+ {\text{h.c.}}\ .~~~~~~~
\end{eqnarray}
Here $M_h$ denotes the SM Higgs mass. 
Note that in order to calculate the branching ratio for muon conversion,
one also has to take into account the hadronic matrix elements of the 
quark and gluon operators which depend on the specific properties of 
the participating nucleus $N$. Using once again the
definitions from~\cite{Beneke:2015lba}, we write  
\begin{eqnarray}
\label{eq:BRmToe}
&& {\rm Br}(\mu N  \to e N)
\nn \\[0.2em] 
&=&
\frac{m_\mu^5}{4 \Gamma_{\text{capture}}}\left| A_R \, \mathcal{D} 
+ 4\left[ \frac{m_\mu m_p}{M_h^2}  \!\left(\tilde C^p_{SL}\!-\!12 \pi \,
\tilde C^p_{L,gg} \right)  \mathcal{S}^p
+ \tilde C^p_{VL}  \, \mathcal{V}^p  
+ \,\{p\rightarrow n\} \right] \right|^{\,2} 
\nn \\[0.4em] && {}
+ \left\lbrace L\leftrightarrow R \right \rbrace\,.
\end{eqnarray}
Conventionally, the branching ratio has been normalised 
to the total capture rate $\Gamma_{\rm capture}$
of the respective nucleus $N$. 
The coefficients $\mathcal{D}$, $\mathcal{S}^{p,n}$, $\mathcal{V}^{p,n}$
in Eq.~\eqref{eq:BRmToe} encode the properties of the target nucleus,
see~\cite{Kitano:2002mt}, where the superscript refers to the proton and neutron
contributions, respectively.
Furthermore, the coefficients $\tilde C_X^p$ are given by  
\begin{align}
& \tilde C^p_{SL} = \sum_{q=u,d,s} c^q_{SL} \, f^p_{q} \,, \qquad 
 \tilde C^p_{VL} =  \sum_{q=u,d} c^q_{VL} \, f^p_{V_q}\,, \qquad
 \tilde C^p_{L,gg} =   c_{gg}^L \, f^p_Q \,,
\end{align}
and analogously for $p\to n$ and $L\to R$. 
Here, the form factors 
$f_q^{p,n}$ and $f_{V_q}^{p,n}$ parametrise the coupling strengths of the 
quark scalar and vector currents of flavour $q$ to nucleons, respectively. 
$f_Q^{p,n}$ represent the scalar couplings of heavy quarks ($c,b,t$).
Finally, the genuine LFV effects are contained in
short-distance Wilson coefficients which in the tree-level
approximation read
\begin{align}
 c^q_{SL}= - \frac{1}{\sqrt{2}m_\mu v } \, \Delta g^{12}_{h \ol \ell \ell} \,, & \qquad 
 c^q_{SR}= - \frac{1}{\sqrt{2}m_\mu v } \,[\Delta g^{\dagger}_{h \ol \ell
     \ell}]^{12}\,,
\end{align}
and
\begin{align}
\label{eq:cCoeffs}
 c^u_{VL}= -\frac{1}{v^2} \left[ \frac12 \Delta g^{12}_{Z \ol \ell _L \ell_L}
\left( 1-\frac{8}{3}s_W^2 \right)\right] 
\,, \hspace{3.2mm} &\quad
c^u_{VR}= \frac{1}{v^2} \left[\frac12 \Delta g^{12}_{Z \ol \ell _R
     \ell_R}\left( 1-\frac{8}{3}s_W^2 \right)\right] 
\,,~~\cr 
 c^d_{VL}= -\frac{1}{v^2} \left[ \frac12 \Delta g^{12}_{Z \ol \ell _L \ell_L}
\left(-1+\frac{4}{3}s_W^2 \right)\right] 
\,,   &\quad
c^d_{VR}= \frac{1}{v^2} \left[\frac12 \Delta g^{12}_{Z \ol \ell
    _R \ell_R}\left(-1+\frac{4}{3}s_W^2 \right)\right] 
\,,~~
\end{align}
as well as~\cite{Shifman:1978zn,Crivellin:2014cta}
\begin{equation}
 c^L_{gg}= - \frac{1}{12 \pi } \sum_{q=c,b,t} c^q_{SL},\quad\qquad
 c^R_{gg}= - \frac{1}{12 \pi } \sum_{q=c,b,t} c^q_{SR} \,.
\end{equation}




\section{Numerical analysis\label{sec:5}}
\cleqn

In the previous section, we have given analytic expressions for all relevant 
LFV observables in terms of the anomalous coupling constants of
the SM particles generated by the Pati-Salam model
at low energy scales. 
These coupling constants depend directly on the $3\times 3$ flavour matrices 
$s_l$ and $t^\prime_l$,
where the latter can be expressed in terms of $s_l$ 
and the effective Yukawa matrix~$Y_\ell$ through the relation in
Eq.~\eqref{eq:Yell-eff}. 
As such, all leptonic branching ratios discussed in Section~\ref{sec:4} 
can be understood as complicated functions of the free model parameters in 
the matrix $s_l$. 
With hardly any restriction on the entries of this matrix, 
the phenomenological analysis naturally lends itself to a
numerical scan over a sizeable and representative part of the 
parameter space.
In the following, we employ two scan strategies. 
For the first strategy, we adopt the same ranges for the
flavour-unspecific input
parameters as in the analysis of quark-flavour 
effects performed in~\cite{Feldmann:2015zwa}, i.e. 
\be
\lambda~\in~ [1.5,3] \ ,\qquad
\tan\beta ~\in~[1,15] \ ,\qquad
M ~\in~ [750,2500]~\text{GeV}\ .
\label{eq:scanranges}
\ee
Furthermore, we choose the entries of the diagonal matrix
$\hat s$, cf. Eq.~\eqref{eq:para-ell}, to lie within the interval  $[\frac{1}{3} \,,\,3] 
\times \frac{\lambda\,v_d}{\sqrt{2}\,\hat M^\ell}$ in order to avoid too much tuning 
between $t'_l$ and $s_l$ in Eq.~\eqref{eq:replacement}. 
The mixing angles and phases in 
the unitary matrices $V_s$ and $U_s$ in Eq.~\eqref{eq:para-ell} are allowed to 
be arbitrarily large for this part of the numerical scan.

Our second scan strategy is based on the same ranges for $\lambda$ and 
$\tan\beta$ as in Eq.~\eqref{eq:scanranges}. However, the NP mass scale $M$ is 
fixed at 1~TeV. Adopting the standard CKM convention to parameterise the unitary
matrices $V_s$ and $U_s$, we then investigate the dependence of the
various LFV observables on the associated mixing angles. 
To this end, we generate two datasets:
one dataset (blue or dark grey points) with arbitrary mixing angles, and another
dataset (orange or light grey points) where the mixing angles are restricted to
be small, i.e.\ in the range $ [0,\pi/6]$.

In practice, we first randomly generate entries for the matrix $s_l$,
together with the parameters 
$\lambda$, $\tan\beta$ and $M$ within the above-mentioned ranges. 
For each set of numerical input parameters we compute the flavour matrix 
$t^\prime_l$ through Eq.~\eqref{eq:Yell-eff} so that the measured charged-lepton 
masses are correctly reproduced. 
Each scan comprises $10^5$ model points. For each model point, we calculate
the corresponding anomalous couplings, and from these the rates for 
$\mu \to e \gamma$, $\mu \to 3 e$, $\mu \to e$ conversion as well as
for the corresponding tau decays.
For muon conversion, we use the same low-energy parameters as 
in~\cite{Beneke:2015lba}. In particular, this implies that 
the muon conversion is computed for a gold target nucleus. We note
that the future experiment DeeMe~\cite{Aoki:2012zza} uses a silicon target,
while Mu2E~\cite{Carey:2008zz} and COMET~\cite{Cui:2009zz} propose aluminium 
targets. Following the argumentation of~\cite{Beneke:2015lba}, 
the projected upper limit for Mu2E of $6\times 10^{-17}$ for
aluminium can be translated to a limit of about $10^{-16}$ for gold targets.

\begin{table}[t!!!p]
\centerline{
\begin{tabular}{|c|c|c|}\hline
Process & Current limit & Future limit \\\hline
Br$(\mu\to e\gamma)$ & 
$ 4.2 \times 10^{-13}$ \cite{TheMEG:2016wtm} & 
$ 6 \times 10^{-14}$ \cite{Baldini:2013ke} \\\hline
Br$(\mu\to 3e)$ & 
$ 1 \times 10^{-12}$ \cite{Bellgardt:1987du} & 
$ 1 \times 10^{-16}$ \cite{Berger:2014vba} \\\hline
Br$^{\rm Au}(\mu N \to e N)$ & 
$ 7 \times 10^{-13}$ \cite{Bertl:2006up} & 
$ 1\times 10^{-16}$ \cite{Carey:2008zz} \\\hline
\end{tabular}}
\caption{Current and future experimental limits on the LFV branching ratios.}
\label{tab:limits}
\end{table}


\subsection{Muonic decays}

In order to illustrate the results of our parameter scan, 
we show the branching ratios of $\mu \to e \gamma$, $\mu \to 3 e$ 
and $\mu \to e$ conversion in form of two-dimensional scatter plots.
These indicate the typical range of the branching ratios and the correlation
between different observables. 
We stress that the point density in these scatter plots depends on the particular 
generation of random numbers, and \emph{does not} 
reflect a probability distribution of the respective observable 
in the considered model.

\begin{figure}[p!!!!tbh]
\begin{center}
   \includegraphics[height=0.27\textheight]{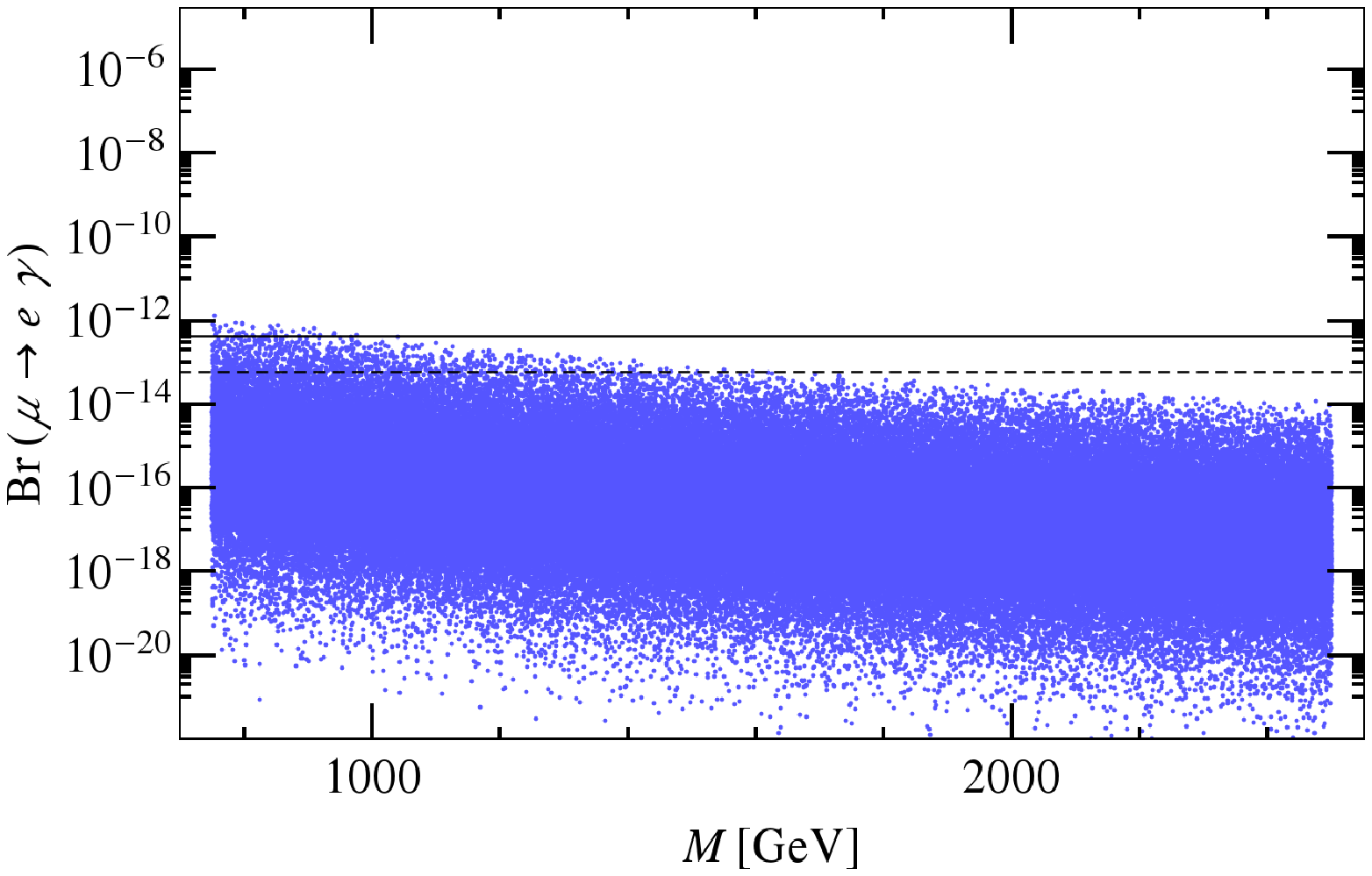}
\\[0.5em]
   \includegraphics[height=0.27\textheight]{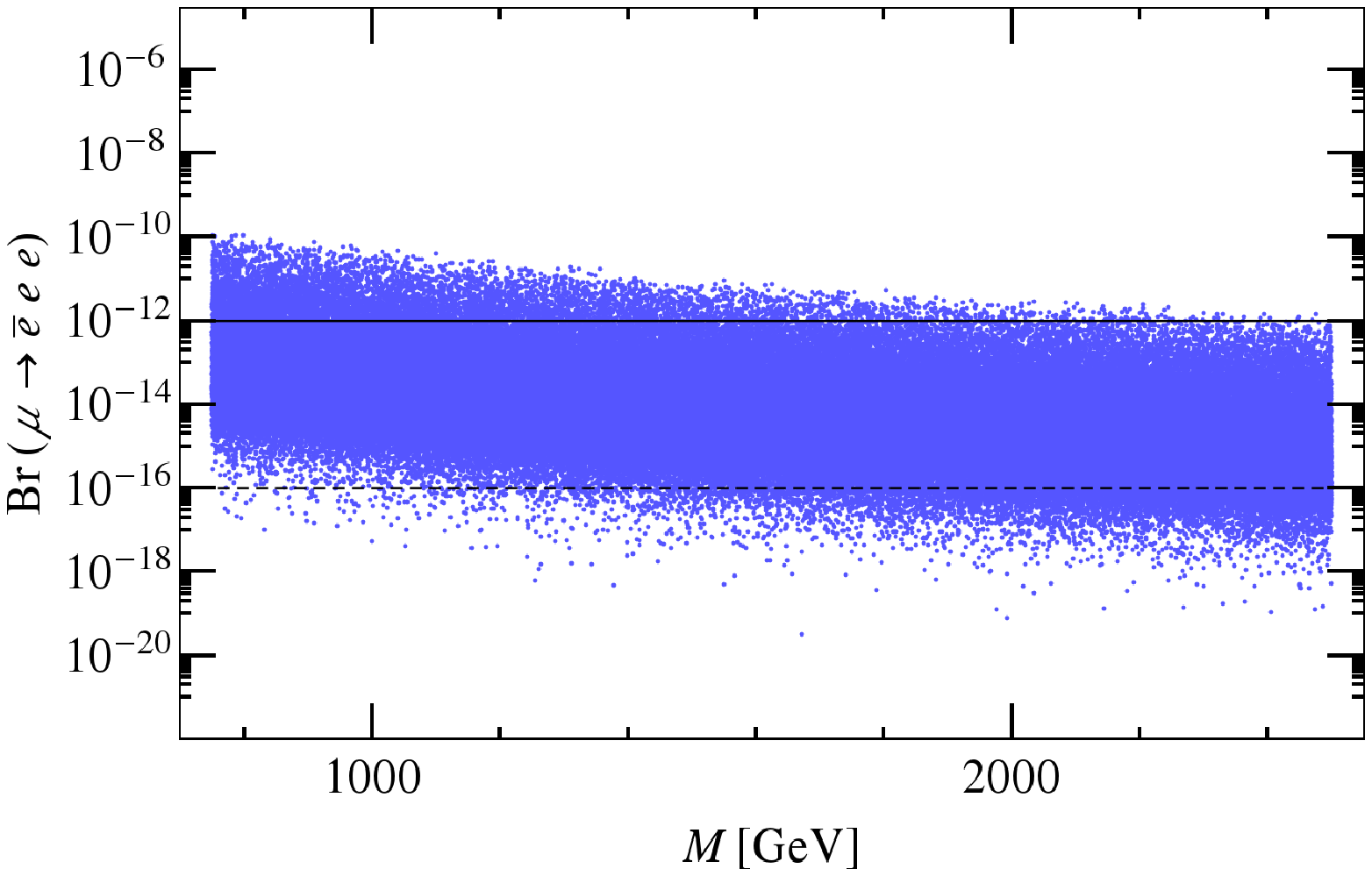}
\\[0.5em]
   \includegraphics[height=0.27\textheight]{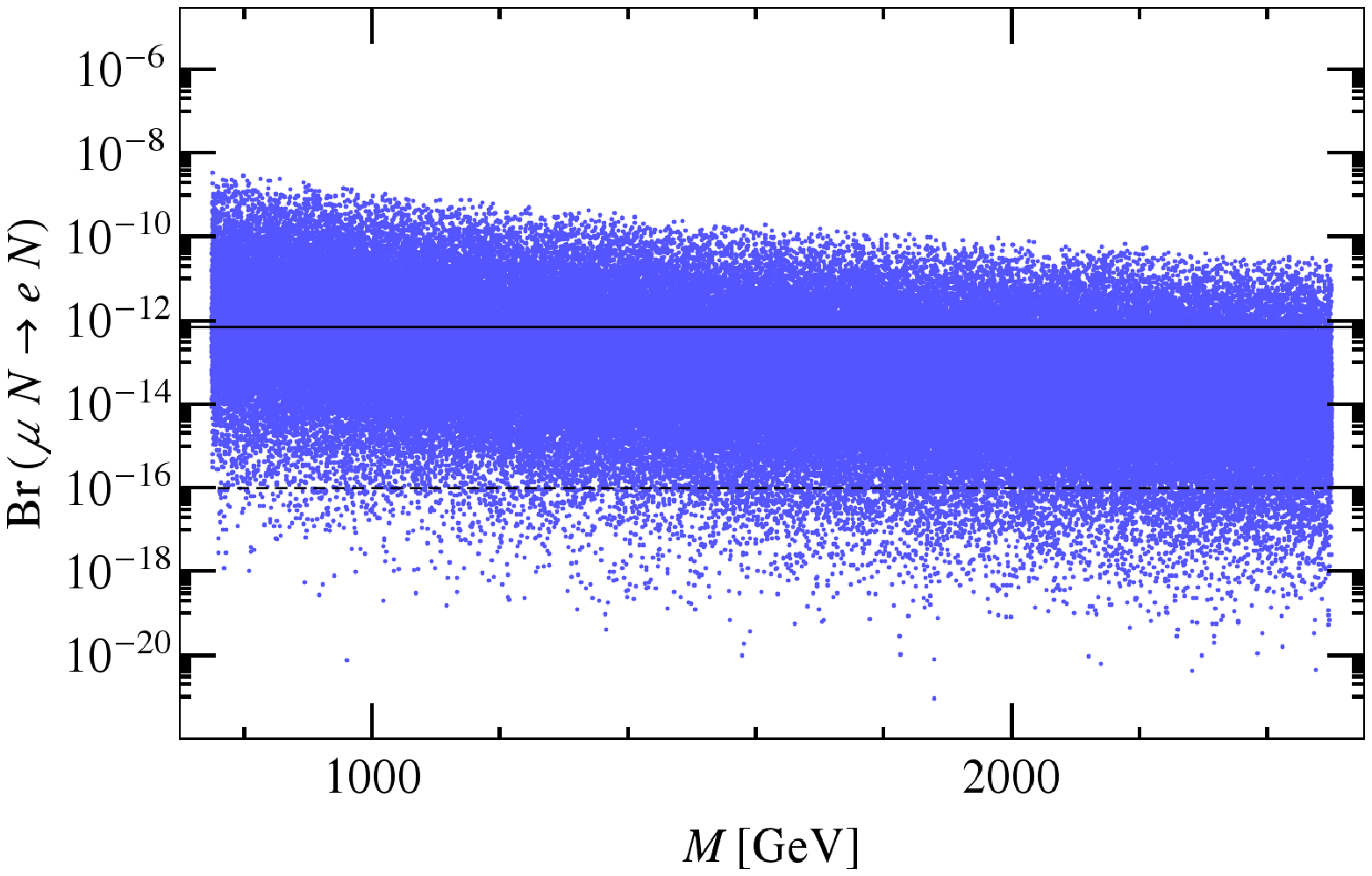}
\end{center}
\caption{\label{fig:LVFvsM} Scatter plots of the branching ratios for $\mu\to e
  \gamma$  (top), $\mu\to 3e $ (centre) and $\mu \to
  e$ conversion (bottom) as a function of the NP scale $M$. The solid
  (dashed) lines indicate the current (future) experimental limits.}
  \vfill 
\end{figure}

Fig.~\ref{fig:LVFvsM} shows the values of the branching ratios for $\mu\to e 
\gamma$ , $\mu \to 3 e$ and muon conversion as a function of the
NP scale~$M$. Each plot contains the current and future experimental
upper limits, see Table~\ref{tab:limits},
visualised  by solid and dashed lines, respectively. 
All three plots show the expected scaling of the branching fractions 
with the small parameter $\epsilon_d^4\propto 1/M^4$. 
We observe that the resulting branching ratios for the decay $\mu \to e \gamma $ are 
largely compatible with the experimental upper bound, even after taking into
account the proposed future upgrade of the MEG experiment~\cite{Baldini:2013ke}. 
The small values for the $\mu \to e\gamma$ branching ratio  
can be explained by the one-loop  suppression of the effective Wilson coefficients 
relevant to this decay in our model. 
In contrast,
the $\mu \to 3 e$ decay and $\mu\to e$ conversion are induced by tree-level 
processes which lead to larger effects, and thus the present and future 
experimental bounds cut stronger into the NP parameter space.
Especially in the case of muon conversion, with the anticipated future 
sensitivity, our model could lead to clear LFV signals.

\begin{figure}[p!!!t]
 \begin{center}
   \includegraphics[height=0.27\textheight]{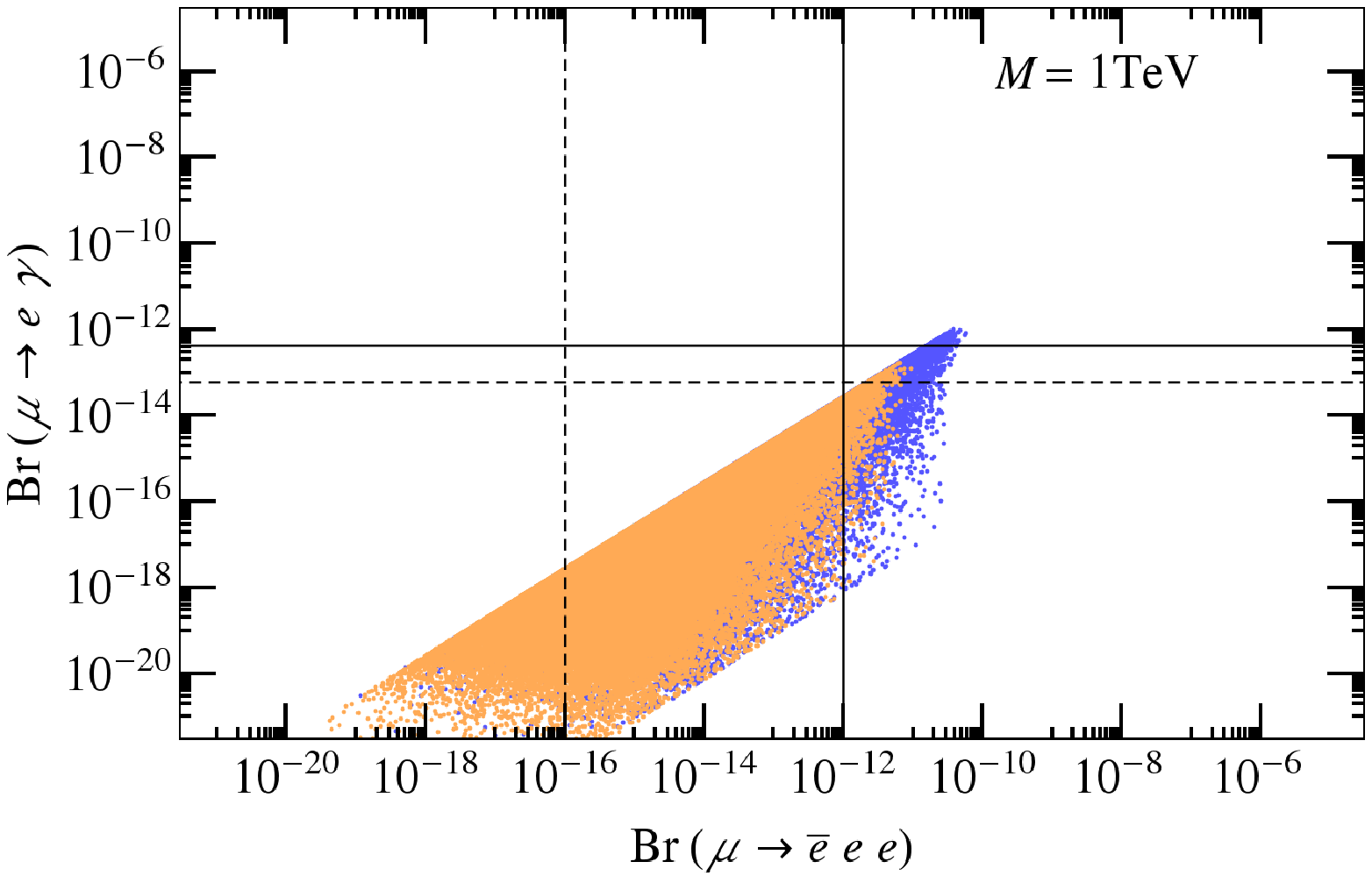}
   \\[0.5em]
   \includegraphics[height=0.27\textheight]{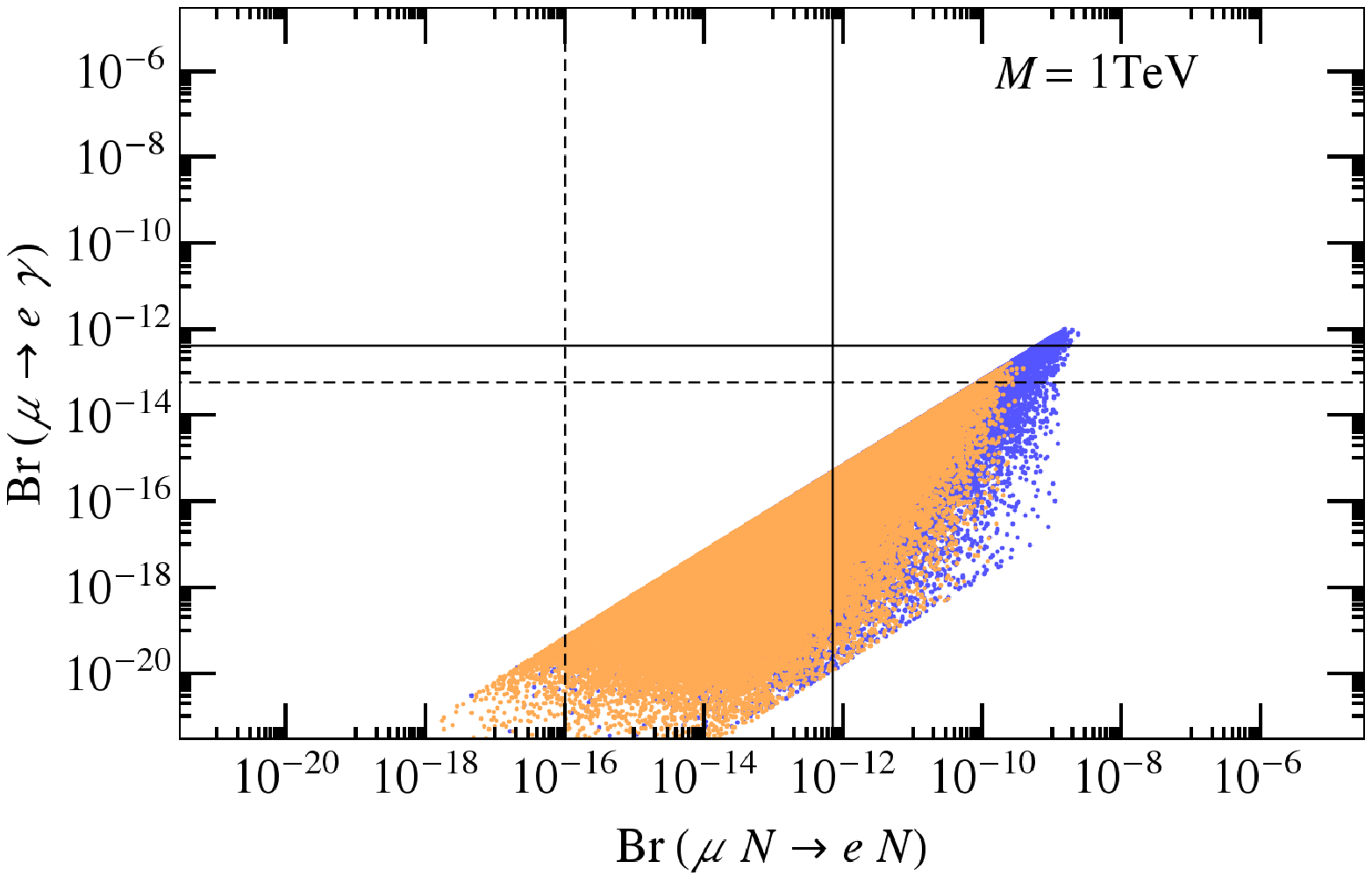}
\\[0.5em]
   \includegraphics[height=0.27\textheight]{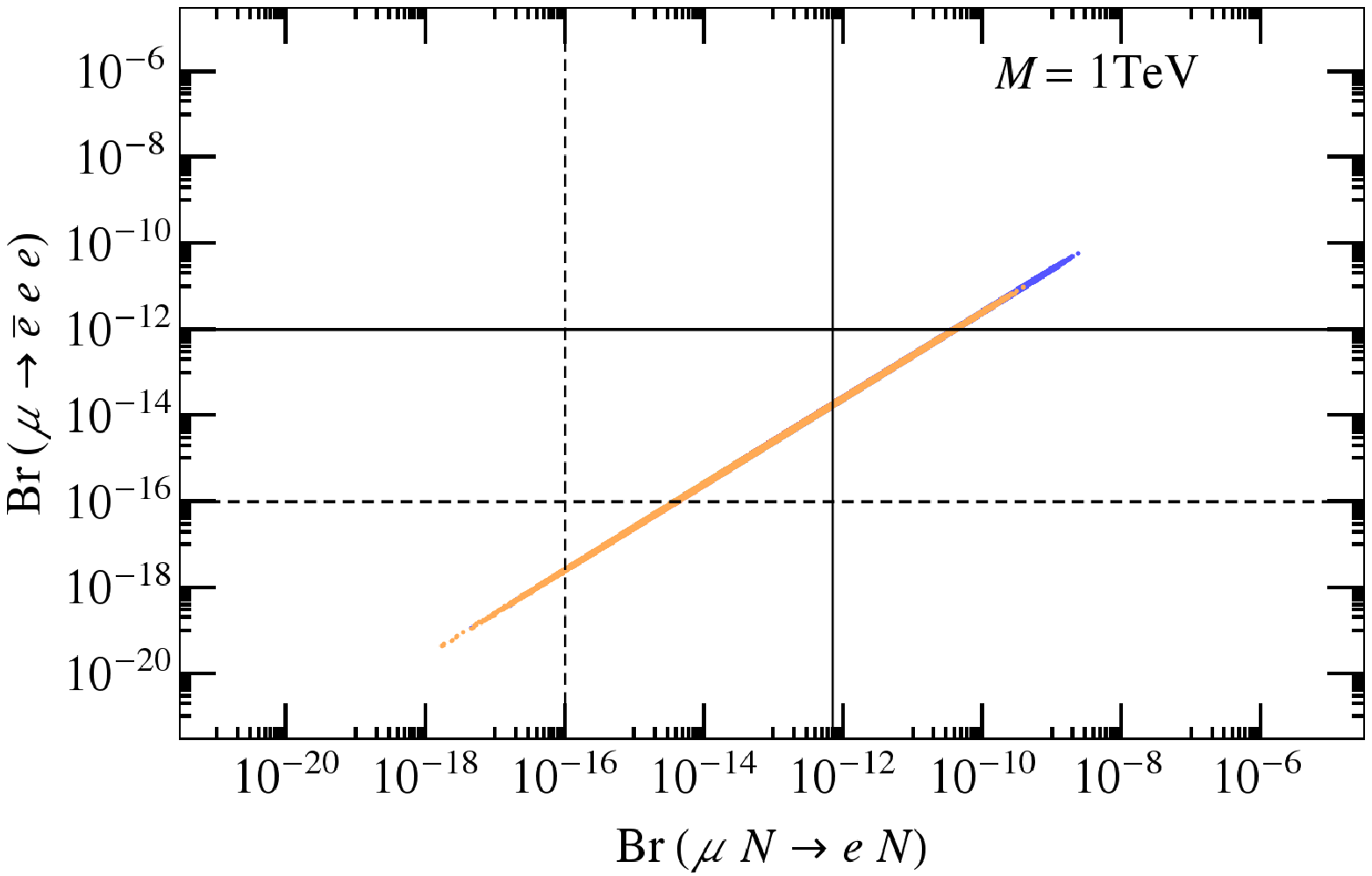}
\end{center}
\caption{\label{fig:MinimalMToECorrelation}  
Correlation between the branching ratios for $\mu\to e \gamma$ vs.\ $\mu \to 3e$ 
(top),   $\mu\to e \gamma$ vs.\ $ \mu N \to  e N$ (centre)
as well as  $\mu\to 3e $ vs.\ $\mu N \to  e N$ (bottom).
Each plot shows the results for a fixed NP scale, $M\!=\!1~\mathrm{TeV}$. 
The scenario with arbitrary 
mixing angles is displayed with blue (dark grey) points, while the scenario 
with mixing angles smaller than $\pi/6$ is shown with orange (light grey) points.
(Orange points are plotted on top of blue points.)}
\end{figure}

To investigate the specific role of the mixing angles in the matrices $U_s$ and $V_s$ 
and the correlations between the different LFV branching ratios, we display in 
Fig.~\ref{fig:MinimalMToECorrelation} the correlations between 
$\mu\to e  \gamma$ and $\mu \to 3e$,  
$\mu\to e \gamma$ and $ \mu N \to  e N$, as well as
$\mu \to 3e$ and $\mu N \to  eN$ 
for a fixed NP scale, $M\!=\!1~\mathrm{TeV}$. 
Orange (light grey) points mark models where the mixing angles in Eq.~\eqref{eq:para-ell} are restricted to values smaller than
$\pi/6$, while blue (dark grey) points indicate the scenario with arbitrary mixing angles.
A common feature of all three plots is that the restriction to small mixing angles 
also leads to smaller branching ratios.\footnote{Note that the points 
with small mixing angles conceal most of the model points with arbitrary
mixing angles in the correlation plots.}
This is easily explained by the general increase of the
anomalous coupling constants with larger mixing angles. 
As both $\mu \to 3e $ and $\mu N\to e N$ are generated by
tree-level contributions to the same kind of Wilson coefficients, 
both processes are strongly correlated, a
feature also known from other LFV models where the dipole operators are
suppressed compared to the 4-fermion operators
(see for instance the Randall-Sundrum scenario with 
the Higgs field localised on the UV-brane as discussed 
in~\cite{Beneke:2015lba,Agashe:2006iy}). 
On the other hand, the loop-induced decay $\mu \to e \gamma$ shows only weak
correlations with both $\mu \to 3e $ and $\mu N \to e N$.
These generic features of the LFV phenomenology distinguish our model from other 
extensions of the SM such as Little Higgs models 
(see e.g.\ \cite{Blanke:2007db,otherLHT2,Blanke:2009am,otherLHT1}),
supersymmetric scenarios
(see e.g.\ \cite{Ellis:2002fe,Brignole:2004ah,Paradisi:2005tk,Paradisi:2006jp,Calibbi:2009wk,Feruglio:2009hu,Dimou:2015yng,Dimou:2015cmw}), 
left-right symmetric models
(see e.g.\ \cite{Dev:2013oxa})
or models with a fourth fermion generation (see e.g.\ \cite{Buras:2010cp}).

\subsection{LFV tau decays and the electron EDM}

\begin{figure}[t]
 \begin{center}
   \includegraphics[height=0.27\textheight]{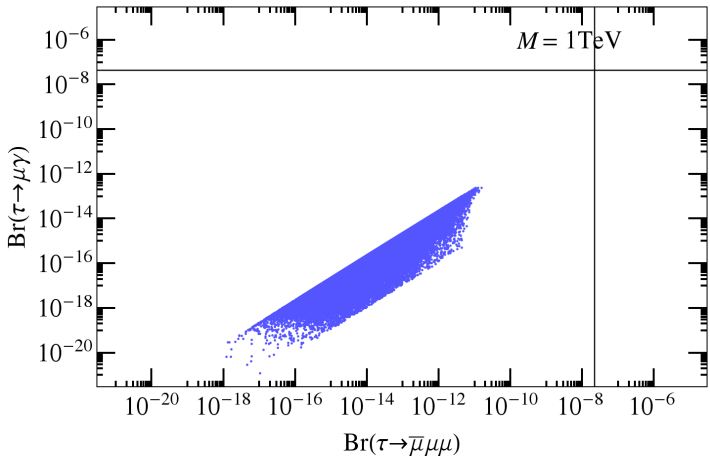}
\end{center}
\caption{\label{fig:tauCorrelation}  
Correlation between the branching ratios of $\tau\to \mu \gamma$ and 
$\tau \to 3\mu$ for a fixed NP scale, $M\!=\!1~\mathrm{TeV}$.}
\end{figure}

In addition to the muon sector, the decays of the tau lepton offer another opportunity 
to observe LFV. However the $\tau$ is not suitable for low-energy experiments 
due to its high mass and short lifetime. The best bounds on processes like 
$\tau \to \mu \gamma$ have been obtained at the BaBar~\cite{Aubert:2009ag}, 
Belle~\cite{Hayasaka:2010np} and LHCb~\cite{Aaij:2013fia} experiments.
The solid lines in Fig.~\ref{fig:tauCorrelation} show the current experimental limits,
\bea
\mathrm{Br}(\tau\to \mu \gamma) &<& 4.4 \times 10^{-8}  
~\text{\cite{Aubert:2009ag}}
\ ,\\
\mathrm{Br}(\tau \to 3\mu) &<& 2.1 \times 10^{-8} 
~\text{\cite{Hayasaka:2010np}}
\ ,
\eea
which can be compared to the predictions of our model for the
corresponding branching ratios with $M\!=\!1~\mathrm{TeV}$, displayed by the blue
points (again referring to the scenario with arbitrary mixing angles for $V_s$
and $U_s$ in Eq.~\eqref{eq:para-ell}). 
Compared to the corresponding muonic case, the plot in
Fig.~\ref{fig:tauCorrelation} shows a similar correlation. However, the
predicted values for the branching ratios are at least three magnitudes smaller
than the best upper limits and will not be accessible in the near future.

We conclude the discussion of the lepton-flavour phenomenology in our
model by considering the strong experimental constraints on the electron 
electric dipole moment (EDM). 
Using the above low-energy Lagrangian, the electron EDM can be directly 
calculated via~(see e.g.\ \cite{Beneke:2015lba})
\begin{align}
 d_e \, = \, m_e \, i\, (A_R-A_L).
\end{align}
However, up to terms proportional to $\mathcal{O}(\epsilon_d^2)$, the anomalous 
coupling matrices, generated by the Pati-Salam model at low energies, are hermitian. 
Thus, the diagonal elements are real, and our setup does not generate any
contributions to the EDM at one-loop level, as long as we restrict ourselves to
the insertion of only one anomalous coupling, i.e.\ the effect of 
dimension-six operators. 
This raises the question whether dimension-eight terms could provide significant
contributions to the electron EDM. To estimate the effects of such dimension-eight
operators, we have calculated diagrams of the type shown in
Fig.~\ref{fig_ARL} with two anomalous gauge boson couplings instead of
one. 
Formally, such diagrams correspond to dimension-eight contributions where the
imaginary part does not vanish for two external electron states with, for instance, a 
$Z$~boson exchanged inside the loop. We find that such a diagram contributes
to the electron EDM only by a numerically negligible amount which, 
for all points of our dataset, 
is at least two orders of magnitude smaller than the measured upper
limit~\cite{Baron:2013eja} 
\begin{equation}
\label{eldipoleconstraint}
|d_e| < 8.7 \cdot 10^{-29} \,e \;\mbox{cm} \ .
\end{equation} 
Assuming normal power counting, we thus conjecture that the contributions of 
other dimension-eight operators to the electron EDM will be parametrically
suppressed in a similar manner.  




\section{Conclusions\label{sec:6}}
\cleqn

We have shown how the Pati-Salam model with gauged 
$SU(3)_I \times SU(3)_{II}$ flavour symmetry constructed 
in~\cite{Feldmann:2015zwa} can be extended to yield a realistic description of
the charged-lepton flavour sector. It requires a non-trivial extension of the
flavour symmetry breaking scalar sector, involving VEVs of flavon fields which
transform non-trivially with respect to the Pati-Salam symmetry group. The
model features two heavy fermionic partners for each SM fermion, whose mixing
with the charged leptons gives rise to anomalous couplings with the SM gauge
bosons as well as the Higgs boson. Expressing the low-energy effective
Lagrangian in terms of these anomalous couplings, we have determined the
branching ratios for $\mu\to e\gamma$, $\mu\to 3e$, $\mu \to e$ conversion as
well as for the corresponding tau decays both analytically and numerically.

Our phenomenological analysis shows that, for the bulk of parameter
space of the model, we do not expect to see any experimental evidence for the
decay $\mu \to e\gamma$ in the near future. This is similar to other LFV
models in which the decay is only induced at the one-loop level. 
On the other hand,  $\mu\to 3 e$ decays and $\mu\to e$ conversion in nuclei are
induced by tree-level processes in our setup. This entails branching ratios
which are accessible with future experimental sensitivity, provided the new-physics
scale associated to the heavy fermions in our model is of the order of a
few~TeV. 
Moreover, for the parameter range assumed in our scans, we found that both
branching ratios are highly correlated.

In the case of lepton-flavour violating $\tau$ decays, we found that the branching
ratios predicted within our model turn out to be orders of magnitude smaller
than the corresponding current experimental limits. Similarly,
contributions to the electron EDM are parametrically suppressed and safely
below the present experimental bound.

In conclusion, our particular model setup, which combines 
the idea of grand unification in the gauge sector and 
flavour symmetry breaking transferred by new heavy 
vector-like fermions, leads to small flavour-violating
effects in the charged-lepton sector, without imposing 
the concept of minimal flavour violation in the 
technical sense (as defined in~\cite{Cirigliano:2005ck},
see also the critical discussion in~\cite{Davidson:2006bd}). 
In particular, the neutrino sector with the phenomenology
of the PMNS matrix is completely decoupled from 
LFV in the charged-lepton sector.




\section*{Acknowledgements}

This work is supported by the Deutsche Forschungsgemeinschaft
(DFG) within the Research Unit FOR 1873
(``Quark Flavour Physics and Effective Field Theories'').  

\section*{Appendix}
\begin{appendix}




\section{\label{app:basis-transformations}Details of required basis transformations}
\cleqn


\subsection{\label{app:diagonalize}Diagonalising the charged-lepton mass matrix}

In Section~\ref{sec:diag-ell}, we have only sketched the diagonalisation
of the $9\times 9$ mass matrix of the charged leptons as defined in
Eq.~\eqref{eq:Me}. Here, we fill in the gaps by explicitly defining the
individual steps [labelled by a subscript $(i)$] of the sequence of basis
transformations. We follow closely the discussion in Appendix~C
of~\cite{Feldmann:2015zwa}. 

\begin{enumerate}

\item Basis with diagonal $Y_\ell = \hat Y_\ell$:

We begin with $\m M^\ell$ of Eq.~\eqref{eq:Me} in a basis where
$Y_\ell$, as defined in Eq.~\eqref{eq:Yell-eff}, is diagonal. Generally,
neither $s_l$ nor $s_l-t'_l$ will be diagonal in that basis. We therefore rewrite
these two matrices using the singular value decomposition of
Eq.~\eqref{eq:para-ell}.

\item Diagonalising $s_l$ and $s_l-t'_l$:

Next, we apply the basis transformation
\be\label{eq:step2}
{\ol \Psi^\ell_L}_{(2)} ~=~ {\ol \Psi^\ell_L}_{(1)}
~\mathrm{diag}\,(U_s^\dagger,V_t^\dagger,V_s^\dagger) \ , \qquad
{\Psi^\ell_R}_{(2)} ~=~
~\mathrm{diag}\,(V_t,U_t,U_s)  {\Psi^\ell_R}_{(1)} \ ,
\ee
so that 
\be
\mathcal M^\ell_{(2)} ~=~
\begin{pmatrix} 
0 & U^\dagger \,  \lambda \epsilon_d & \id \\
\id  &\hat t  & 0  \\
V^\dagger\,  \lambda  \epsilon_d  & 0 &  \hat s
\end{pmatrix} M \ , \label{eq:Me-rot2}
\ee
with
\be
U^\dagger = U _sU_t^\dagger \ , \qquad
V^\dagger = V_sV_t^\dagger \ .
\ee
The matrices $\hat s$ and $\hat t$ are the diagonal versions of $s_l$ and
$s_l-t'_l$, see Eq.~\eqref{eq:para-ell}.
The structure of Eq.~\eqref{eq:Me-rot2} is identical to the one of Eq.~(C.2)
in~\cite{Feldmann:2015zwa}. Therefore, we can simply follow the sequence of
basis transformations described in Appendix~C of~\cite{Feldmann:2015zwa}.

\item Diagonalising $\m M^\ell_{(2)}$ for $\epsilon_d=0$:

With vanishing $\epsilon_d$, the three generations do not mix with each
other. It is therefore straightforward to diagonalise $\m M^\ell_{(2)}$ in
this limit by introducing the following matrices of cosines and sines
\be
\hat c_x = \text{diag} (c_x^1,c_x^2,c_x^3) \ , \qquad
\hat s_x = \text{diag} (s_x^1,s_x^2,s_x^3) \ ,
\ee
where 
\be
    c^i_x = \frac{\hat{x}_i}{\sqrt{1+{\hat x}_i^2}}  \ ,  \qquad
   s^i_x = \frac{1}{\sqrt{1+{\hat x}_i^2}} \ ,\label{eq:cossin}
\ee
and $x=s,t$. Defining the basis transformation
\be
{\overline \Psi^\ell_L}_{(3)} \approx {\overline \Psi^\ell_L}_{(2)}\!
\begin{pmatrix} 
\hat c_s  & 0 &  \hat s_s \\
0&\id&0 \\
-\hat s_s & 0 & \hat c_s
\end{pmatrix}, \label{eq:step3a}
\qquad
{\Psi^\ell_R}_{(3)} \approx 
\begin{pmatrix}
\hat c_t& -\hat s_t & 0 \\
\hat s_t & \hat c_t& 0\\
0&0&\id 
\end{pmatrix} \!
{\Psi^\ell_R}_{(2)} \ ,
\ee
diagonalises $\m M^\ell_{(2)}$ in the limit where $\epsilon_d=0$. 
Reinstating non-vanishing $\epsilon_d$ yields the mass
matrix $\m M^\ell_{(3)}$ which is of the form as given in
Eq.~\eqref{eq:Mell-basis3}. The exact form of the $3\times 3$ submatrices
$a$, $b$, $c$, $d$, $\hat e$ and $\hat f$ was derived
in~\cite{Feldmann:2015zwa}. In this work, we can simplify these expressions
thanks to the small Yukawa coupling of the tau lepton, and hence 
$\hat x_i\gg 1$. Expanding to second order in~$\hat x_i^{-1}$ we have
\be\label{eq:c-s-expansion}
\hat c_x ~\approx ~\id -\frac{1}{2} \hat x^{-2} \ , \qquad
\hat s_x ~\approx~  \hat x^{-1} \ ,
\ee
and with it
\bea
a&\approx & - (\hat  s^{-1} V^\dagger  +  U^\dagger \hat t^{-1}) \,\lambda\ ,\label{eq:a}
\\
b&\approx &   \left(
U^\dagger 
- \hat s^{-1} V^\dagger \hat t^{-1}  
-\mbox{$\frac{1}{2}$} \hat   s^{-2} U^\dagger 
- \mbox{$\frac{1}{2}$} U^\dagger \hat t^{-2}
\right)
\, \lambda\ ,\label{eq:b}\\
c&\approx &   \left(
V^\dagger 
- \hat s^{-1} U^\dagger \hat t^{-1}  
-\mbox{$\frac{1}{2}$} \hat   s^{-2} V^\dagger 
- \mbox{$\frac{1}{2}$} V^\dagger \hat t^{-2}
\right)
\, \lambda\ ,\label{eq:c}\\
d&\approx & ( V^\dagger \hat  t^{-1} +  \hat s^{-1} U^\dagger ) \,\lambda\ ,\label{eq:d}\\
\hat e &\approx& \hat t \ ,\label{eq:e}\\
\hat f &\approx& \hat s \ .\label{eq:f}
\eea

\item Block-diagonalising $\m M^\ell_{(3)}$ up to order $\epsilon_d^2$:

The matrix in Eq.~\eqref{eq:Mell-basis3} can be block-diagonalised (to second order
in $\epsilon_d$) by
\bea \label{eq:basis4a}
{\overline \Psi^\ell_L}_{(4)} &\equiv & {\overline \Psi^\ell_L}_{(3)}
 \left[\mathcal R_{12} (\xi^{\ell}_{12})\right]^\dagger
\left[\mathcal R_{23} (\xi^{\ell}_{23})\right]^\dagger
\left[\mathcal R_{13} (\xi^{\ell}_{13})\right]^\dagger
 \ , \\[1mm] \label{eq:basis4b}
{\Psi^\ell_R}_{(4)} &\equiv& 
\left[ \mathcal R_{12} (\zeta^{\ell}_{12})\right]\,
\left[ \mathcal R_{23} (\zeta^{\ell}_{23})\right]\,
\left[ \mathcal R_{13} (\zeta^{\ell}_{13})\right]
 \, {\Psi^\ell_R}_{(3)} \ .
\eea
Here $ \mathcal{R}_{\alpha\beta}(\xi)$ denotes a ``rotation in the
$\alpha$-$\beta$ plane'', expanded to second order in~$\xi$. For example, 
\begin{align}
  \mathcal{R}_{12}(\xi) = \left(\begin{array}{ccc} \id - \tfrac12\,
\xi\,\xi^\dagger & -\xi & 0\\ \xi^\dagger & \id - \tfrac12\,\xi^\dagger\,\xi & 0 \\
0&0&\id \end{array} 
\right) \, ,
\end{align}
while the other two rotations are identical up to obvious permutations of rows
and columns. In terms of the parameters of Eq.~\eqref{eq:Mell-basis3}, the
$3\times 3$ matrices $\xi$ and $\zeta$ have already been derived in~\cite{Feldmann:2015zwa},
\begin{align}
{\xi}^{\ell}_{12} &= b \hat e^{-1} \, \epsilon_d \ ,
&\left[{\xi}^{\ell}_{23}\right]_{ij} &= 
\frac{-\hat e_i  d^\dagger_{ij}}
{\phantom(\!\!\hat e_i\phantom)\!\!^2 -\phantom(\!\!\hat f_j\phantom)\!\!^2} 
\,\epsilon_d \ ,
&{\xi}^{\ell}_{13} &= a c^\dagger  \hat f^{-2}\,\epsilon_d^2 \ ,
\\
{\zeta}^{\ell}_{12} &= a^\dagger b  \hat e^{-2} \, \epsilon_d^2 \ ,
&\left[{\zeta}^{\ell}_{23}\right]_{ij} &= 
\frac{-  d^\dagger_{ij}\hat f_j}
{\phantom(\!\!\hat e_i\phantom)\!\!^2 -\phantom(\!\!\hat f_j\phantom)\!\!^2} 
\,\epsilon_d \ ,
&{\zeta}^{\ell}_{13} &= c^\dagger  \hat f^{-1}\,\epsilon_d \ .
\end{align} 
Inserting the expressions of Eqs.~(\ref{eq:a}--\ref{eq:f}) and keeping only
terms up to second order in $\hat s^{-1}$ and~$\hat t^{-1}$, we find for
the charged leptons 
\begin{align}\label{eq:xiellsimpl}
{\xi}^{\ell}_{12} &\approx U^\dagger  \hat t^{-1} \, \lambda\,\epsilon_d \ ,
&\left[{\xi}^{\ell}_{23}\right]_{ij} &\approx 
\frac{-  V_{ij}  - \hat t_i  \, U_{ij}\, \hat s_j^{-1}}
{\phantom(\!\!\hat t_i\phantom)\!\!^2 -\phantom(\!\!\hat s_j\phantom)\!\!^2} 
\,\lambda\,\epsilon_d \ ,
&{\xi}^{\ell}_{13} &\approx  0 \ ,
\\\label{eq:zetaellsimpl}
{\zeta}^{\ell}_{12} &\approx  0 \ ,
&\left[{\zeta}^{\ell}_{23}\right]_{ij} &\approx  
\frac{
-  \hat t^{-1}_i \, V_{ij}\,\hat s_j  -   U_{ij}}
{\phantom(\!\!\hat t_i\phantom)\!\!^2 -\phantom(\!\!\hat s_j\phantom)\!\!^2} 
\,\lambda\,\epsilon_d \ ,
&{\zeta}^{\ell}_{13} &\approx  V  \hat s^{-1}\,\lambda\,\epsilon_d \ .
\end{align} 
With this result, the basis change in Eqs.~(\ref{eq:basis4a},\ref{eq:basis4b})
simplifies to
\bea \nonumber
 \left[\mathcal R_{12} (\xi^{\ell}_{12})\right]^\dagger
\left[\mathcal R_{23} (\xi^{\ell}_{23})\right]^\dagger &\!\!\approx\!\!&
 \begin{pmatrix}
\id - \tfrac12\, \xi^\ell_{12}\,{\xi^\ell_{12}}^\dagger & \xi^\ell_{12} & 0\\ -{\xi^\ell_{12}}^\dagger &\!\!\! \id - \tfrac12\,{\xi^\ell_{12}}^\dagger\,\xi^\ell_{12}\!\! & 0 \\
0&0&\id 
\end{pmatrix}\!
\begin{pmatrix}
\id&0&0\\
0 & \id &\xi^\ell_{23}  \\
0& -{\xi^\ell_{23}}^\dagger & \id 
\end{pmatrix}~~~~~\\
&\!\!\approx\!\!&
\begin{pmatrix}
\id - \tfrac12\, \xi^\ell_{12}\,{\xi^\ell_{12}}^\dagger & \xi^\ell_{12} & 0\\ -{\xi^\ell_{12}}^\dagger & \id - \tfrac12\,{\xi^\ell_{12}}^\dagger\,\xi^\ell_{12} & \xi^\ell_{23} \\
0& -{\xi^\ell_{23}}^\dagger&\id 
\end{pmatrix},
\label{eq:step4a}
\\[3mm] \nonumber
\left[ \mathcal R_{23} (\zeta^{\ell}_{23})\right]\,
\left[ \mathcal R_{13} (\zeta^{\ell}_{13})\right]
 &\!\!\approx\!\!&
\begin{pmatrix}
\id&0&0\\
0 & \id &-\zeta^\ell_{23}  \\
0& {\zeta^\ell_{23}}^\dagger & \id 
\end{pmatrix}\!
\begin{pmatrix}
\id - \tfrac12\, \zeta^\ell_{13}\,{\zeta^\ell_{13}}^\dagger \!&0& -\zeta^\ell_{13} \\ 
0&\id&0\\
{\zeta^\ell_{13}}^\dagger &0&
\!  \id - \tfrac12\,{\zeta^\ell_{13}}^\dagger\,\zeta^\ell_{13}
\end{pmatrix}~~~~~\\
&\!\!\approx\!\!&
 \begin{pmatrix}
\id - \tfrac12\, \zeta^\ell_{13}\,{\zeta^\ell_{13}}^\dagger &0& -\zeta^\ell_{13} \\ 
0&\id&-\zeta^\ell_{23} \\
{\zeta^\ell_{13}}^\dagger &{\zeta^\ell_{23}}^\dagger &
  \id - \tfrac12\,{\zeta^\ell_{13}}^\dagger\,\zeta^\ell_{13}
 \end{pmatrix} .\label{eq:step4b}
\eea
The resulting mass matrix $\m M^\ell_{(4)}$  is given explicitly in
Eq.~\eqref{eq:Mell-basis4}.

\item The approximate mass basis:

The final step in our sequence of basis transformations diagonalises the upper
left block $av_d/\sqrt{2}$ of $\m M^\ell_{(4)}$. As discussed in
Section~\ref{sec:diag-ell}, this requires
\bea \label{eq:basis5}
{\overline \Psi^\ell_L}_{(5)} =  {\overline \Psi^\ell_L}_{(4)}
 \mathrm{diag}\,(U_s,\id,\id) \ ,\qquad
{\Psi^\ell_R}_{(5)} = 
 \mathrm{diag}\,(V_t^\dagger,\id,\id) 
 \, {\Psi^\ell_R}_{(4)} \ .
\eea

\end{enumerate}


\subsection{\label{app:majorana}The neutral-lepton mass matrix}

Considering the neutral fermions of the Pati-Salam model, we do not aim at a
full diagonalisation of the $18 \times 18$ Majorana mass matrix of
Eq.~\eqref{eq:18x18}. Yet, we must identify the correct light neutrino mass
eigenstates which become massless in the limit of $\epsilon_u=0$. 
Working in this limit, we apply a basis transformations on
$\Psi_{\mathrm{Maj}}$ of Eq.~\eqref{eq:Maj-vector} such that
\bea
{\overline \Psi^\nu_L}_{(3)}^T &=& 
\begin{pmatrix} 
\hat c_s  & 0 & - \hat s_s \\
0&\id&0 \\
\hat s_s & 0 & \hat c_s
\end{pmatrix}
\begin{pmatrix}
U_s^\ast&0&0\\
0&\id&0\\
0&0&V_s^\ast
\end{pmatrix}
{\overline \Psi^\nu_L}_{(1)}^T
\ ,
\label{eq:majstep3a}
\\[3mm]
{\Psi^\nu_R}_{(3)} &=& 
\begin{pmatrix}
\id&0&0\\
0&\id&0\\
0&0&U_s
\end{pmatrix}
{\Psi^\nu_R}_{(1)}\ .\label{eq:majstep3b}
\eea
Adopting the expansion of Eq.~\eqref{eq:c-s-expansion}, the mass
matrix $\m M^{\mathrm{Maj}}_{(3)}\big|_{\epsilon_u=0}$ takes the form of
Eq.~\eqref{eq:18x18simpl}, in which the three massless left-handed neutrinos
have already been isolated. The fourth step in the sequence of basis
transformations must not be applied to the neutral sector as it would add a
component of the heavy fermions to the massless particles; hence, $\xi^\nu_{ij}=0$. 
Finally, the fifth step only mixes the massless neutral fermions and can be
chosen identical to the corresponding unitary matrix of the charged
leptons. We therefore define
\be\label{eq:majstep5}
{\overline \Psi^\nu_L}^T_{(5)} = \,\mathrm{diag}\,(U_s^T,\id,\id) \,
 {\overline \Psi^\nu_L}^T_{(3)}
   \ .
\ee
We do not specify the transformations of the heavy neutral fermions as they
are practically irrelevant for charged-lepton flavour violating observables
involving only the electron, the muon and the tau lepton.


\subsection{\label{app:gaugekin}Non-standard gauge-kinetic couplings}

In the neutral currents of the charged leptons, deviations from the
Standard Model originate from the terms of Eq.~\eqref{eq:GRVbrokenkinlag}
which are proportional to $\mathcal K'_L$ and $\mathcal K_R$. Applying the
sequence of basis transformations defined in Appendix~\ref{app:diagonalize}  yields
\bea
{\mathcal K^{\prime\,0}_L}_{(5)} &=&
\begin{pmatrix}
U_s^\dagger \xi^\ell_{12} {\xi^\ell_{12}}^\dagger U_s & -U_s^\dagger
\xi^\ell_{12} &0\\
-{\xi^\ell_{12}}^\dagger U_s & 1-{\xi^\ell_{12}}^\dagger \xi^\ell_{12} &
\xi^\ell_{23}\\
0&{\xi^\ell_{23}}^\dagger& 0 
\end{pmatrix} , \\
{\mathcal K^0_R}_{(5)} &=&
\begin{pmatrix}
V_t^\dagger \zeta^\ell_{13} {\zeta^\ell_{13}}^\dagger V_t & 0 & -V^\dagger_t
\zeta^\ell_{13}\\
0&0& -\zeta^\ell_{23} \\
-{\zeta^\ell_{13}}^\dagger V_t & -{\zeta^\ell_{23}}^\dagger & 1- 
{\zeta^\ell_{13}}^\dagger \zeta^\ell_{13}
\end{pmatrix} ,
\eea
with $\xi^\ell_{ij}$ and $\zeta^\ell_{ij}$ given in
Eqs.~(\ref{eq:xiellsimpl},\ref{eq:zetaellsimpl}). 

Turning to the charged current involving the left-handed neutrino, we have not
fully specified the transformation which diagonalises the mass matrix of all neutral
fermions. In particular, we have skipped the diagonalisation of the Majorana
mass matrix of Eq.~\eqref{eq:9x9simpl}. Therefore, we will only determine the
upper left $3\times 3$ block of ${\m K_L^+}_{(5)}$ which describes the
anomalous coupling of the $W$ to the light SM leptons. 
It is easy to see that a change from the original basis of the charged and
neutral leptons to the corresponding third basis, as defined in
Eqs.~(\ref{eq:step2},\ref{eq:step3a},\ref{eq:majstep3a}), does not 
modify the matrix $\mathcal K_L$ at all. Performing the transformations of
Eqs.~(\ref{eq:step4a},\ref{eq:basis5},\ref{eq:majstep5})
 to the final mass basis, one can derive the following form for the upper left
 $3\times 3$ block of ${\m K_L^+}_{(5)}$,
\be
U_s^\dagger \,\left( 
\id - \frac{1}{2} \, \xi^\ell_{12} { \xi^\ell_{12}}^\dagger
\right)\,U_s^{} \ ,
\ee
where $\xi^\ell_{12}$ is given in Eq.~\eqref{eq:xiellsimpl}.

\end{appendix}





\begin{thebibliography}{99}


\bibitem{Aad:2012tfa}
  G.~Aad {\it et al.} [ATLAS Collaboration],
  Phys.\ Lett.\ B {\bf 716} (2012) 1
  [arXiv:1207.7214].

\bibitem{Chatrchyan:2012xdj}
  S.~Chatrchyan {\it et al.} [CMS Collaboration],
  Phys.\ Lett.\ B {\bf 716} (2012) 30
  [arXiv:1207.7235].
 
\bibitem{Buchmuller:1985jz}
  W.~Buchm\"uller and D.~Wyler,
  Nucl.\ Phys.\ B {\bf 268} (1986) 621.
  
\bibitem{Grzadkowski:2010es}
  B.~Grzadkowski, M.~Iskrzynski, M.~Misiak and J.~Rosiek,
  JHEP {\bf 1010} (2010) 085
  [arXiv:1008.4884].

\bibitem{Alonso:2013hga}
  R.~Alonso, E.~E.~Jenkins, A.~V.~Manohar and M.~Trott,
  JHEP {\bf 1404} (2014) 159
  [arXiv:1312.2014].

\bibitem{Altmannshofer:2015zpk}
  W.~Altmannshofer,
  PoS FPCP {\bf 2015} (2015) 001.

\bibitem{Chivukula:1987py}
  R.~S.~Chivukula and H.~Georgi,
  Phys.\ Lett.\ B {\bf 188} (1987) 99.

\bibitem{Hall:1990ac}
  L.~J.~Hall and L.~Randall,
  Phys.\ Rev.\ Lett.\  {\bf 65} (1990) 2939.

\bibitem{Buras:2000dm}
  A.~J.~Buras, P.~Gambino, M.~Gorbahn, S.~J\"ager and L.~Silvestrini,
  Phys.\ Lett.\ B {\bf 500} (2001) 161
  [hep-ph/0007085].
  
\bibitem{D'Ambrosio:2002ex}
  G.~D'Ambrosio, G.~F.~Giudice, G.~Isidori and A.~Strumia,
  Nucl.\ Phys.\ B {\bf 645} (2002) 155
  [hep-ph/0207036].
  
\bibitem{Cirigliano:2005ck}
  V.~Cirigliano, B.~Grinstein, G.~Isidori and M.~B.~Wise,
  Nucl.\ Phys.\ B {\bf 728} (2005) 121
  [hep-ph/0507001].
  
\bibitem{Feldmann:2009dc}
  T.~Feldmann, M.~Jung and T.~Mannel,
  Phys.\ Rev.\ D {\bf 80} (2009) 033003
  [arXiv:0906.1523].
  
\bibitem{Alonso:2011yg}
  R.~Alonso, M.~B.~Gavela, L.~Merlo and S.~Rigolin,
  JHEP {\bf 1107} (2011) 012
  [arXiv:1103.2915].

\bibitem{Nardi:2011st}
  E.~Nardi,
  Phys.\ Rev.\ D {\bf 84} (2011) 036008
  [arXiv:1105.1770].
 
\bibitem{Alonso:2012fy}
  R.~Alonso, M.~B.~Gavela, D.~Hernandez and L.~Merlo,
  Phys.\ Lett.\ B {\bf 715} (2012) 194
  [arXiv:1206.3167].
  
\bibitem{Espinosa:2012uu}
  J.~R.~Espinosa, C.~S.~Fong and E.~Nardi,
  JHEP {\bf 1302} (2013) 137
  [arXiv:1211.6428].
 
\bibitem{Alonso:2013mca}
  R.~Alonso, M.~B.~Gavela, D.~Hernández, L.~Merlo and S.~Rigolin,
  JHEP {\bf 1308} (2013) 069
  [arXiv:1306.5922].
  
\bibitem{Alonso:2013nca}
  R.~Alonso, M.~B.~Gavela, G.~Isidori and L.~Maiani,
  JHEP {\bf 1311} (2013) 187
  [arXiv:1306.5927].
  
\bibitem{Alonso:2013dba}
  R.~Alonso de Pablo,
  arXiv:1307.1904.
  
\bibitem{Fong:2013dnk}
  C.~S.~Fong and E.~Nardi,
  Phys.\ Rev.\ D {\bf 89} (2014)  036008
  [arXiv:1307.4412].
 
\bibitem{Merlo:2015nra}
  L.~Merlo,
  arXiv:1503.03282.
 
\bibitem{Grinstein:2010ve}
  B.~Grinstein, M.~Redi and G.~Villadoro,
  JHEP {\bf 1011} (2010) 067
  [arXiv:1009.2049].

\bibitem{Albrecht:2010xh}
  M.~E.~Albrecht, T.~Feldmann and T.~Mannel,
  JHEP {\bf 1010} (2010) 089
  [arXiv:1002.4798].

\bibitem{D'Agnolo:2012ie}
  R.~T.~D'Agnolo and D.~M.~Straub,
  JHEP {\bf 1205} (2012) 034
  [arXiv:1202.4759].
  
\bibitem{Bishara:2015mha}
  F.~Bishara, A.~Greljo, J.~F.~Kamenik, E.~Stamou and J.~Zupan,
  JHEP {\bf 1512} (2015) 130
  [arXiv:1505.03862].

\bibitem{Feldmann:2010yp}
  T.~Feldmann,
  JHEP {\bf 1104} (2011) 043
  [arXiv:1010.2116].
  
\bibitem{Guadagnoli:2011id}
  D.~Guadagnoli, R.~N.~Mohapatra and I.~Sung,
  JHEP {\bf 1104} (2011) 093
  [arXiv:1103.4170].
  
\bibitem{Mohapatra:2012km}
  R.~N.~Mohapatra,
  AIP Conf.\ Proc.\  {\bf 1467} (2012) 7
  [arXiv:1205.6190].
 
  \bibitem{Feldmann:2015zwa}
  T.~Feldmann, F.~Hartmann, W.~Kilian and C.~Luhn,
  JHEP {\bf 1510} (2015) 160
  [arXiv:1506.00782].

\bibitem{Krnjaic:2012aj}
  G.~Krnjaic and D.~Stolarski,
  JHEP {\bf 1304} (2013) 064
  [arXiv:1212.4860].

\bibitem{Franceschini:2013ne}
  R.~Franceschini and R.~N.~Mohapatra,
  JHEP {\bf 1304} (2013) 098
  [arXiv:1301.3637].
  
\bibitem{Chen:2003zv}
  M.~C.~Chen and K.~T.~Mahanthappa,
  Int.\ J.\ Mod.\ Phys.\ A {\bf 18} (2003) 5819
  [hep-ph/0305088].
 
\bibitem{King:2003jb}
  S.~F.~King,
  Rept.\ Prog.\ Phys.\  {\bf 67} (2004) 107
  [hep-ph/0310204].
  
\bibitem{Mohapatra:2006gs}
  R.~N.~Mohapatra and A.~Y.~Smirnov,
  Ann.\ Rev.\ Nucl.\ Part.\ Sci.\  {\bf 56} (2006) 569
  [hep-ph/0603118].

  \bibitem{Babu:2009fd}
  K.~S.~Babu,
  arXiv:0910.2948.

\bibitem{Altarelli:2010gt}
  G.~Altarelli and F.~Feruglio,
  Rev.\ Mod.\ Phys.\  {\bf 82} (2010) 2701
  [arXiv:1002.0211].
  
\bibitem{King:2013eh}
  S.~F.~King and C.~Luhn,
  Rept.\ Prog.\ Phys.\  {\bf 76} (2013) 056201
  [arXiv:1301.1340].
  
\bibitem{King:2014nza}
  S.~F.~King, A.~Merle, S.~Morisi, Y.~Shimizu and M.~Tanimoto,
  New J.\ Phys.\  {\bf 16} (2014) 045018
  [arXiv:1402.4271].
 
\bibitem{King:2015aea}
  S.~F.~King,
  J.\ Phys.\ G {\bf 42} (2015) 123001
  [arXiv:1510.02091].

\bibitem{Pati:1973uk}
  J.~C.~Pati and A.~Salam,
  Phys.\ Rev.\ D {\bf 8} (1973) 1240.
  
\bibitem{Pati:1974yy}
  J.~C.~Pati and A.~Salam,
  Phys.\ Rev.\ D {\bf 10} (1974) 275
   [Phys.\ Rev.\ D {\bf 11} (1975) 703].
  
\bibitem{Kilian:2006hh}
  W.~Kilian and J.~Reuter,
  Phys.\ Lett.\ B {\bf 642} (2006) 81
  [hep-ph/0606277].
  
\bibitem{Howl:2007hq}
  R.~Howl and S.~F.~King,
  Phys.\ Lett.\ B {\bf 652} (2007) 331
  [arXiv:0705.0301].
 
\bibitem{Calibbi:2009cp}
  L.~Calibbi, L.~Ferretti, A.~Romanino and R.~Ziegler,
  Phys.\ Lett.\ B {\bf 672} (2009) 152
  [arXiv:0812.0342].
  
\bibitem{Braam:2009fi}
  F.~Braam, J.~Reuter and D.~Wiesler,
  AIP Conf.\ Proc.\  {\bf 1200} (2010) 458
  [arXiv:0909.3081].
 
\bibitem{DeRomeri:2011ie}
  V.~De Romeri, M.~Hirsch and M.~Malinsky,
  Phys.\ Rev.\ D {\bf 84} (2011) 053012
  [arXiv:1107.3412].

\bibitem{Arbelaez:2013hr}
  C.~Arbelaez, R.~M.~Fonseca, M.~Hirsch and J.~C.~Rom\~ao,
  Phys.\ Rev.\ D {\bf 87} (2013)  075010
  [arXiv:1301.6085].
 
\bibitem{Arbelaez:2013nga}
  C.~Arbelaez, M.~Hirsch, M.~Malinsky and J.~C.~Rom\~ao,
  Phys.\ Rev.\ D {\bf 89} (2014) 035002
  [arXiv:1311.3228].
 
\bibitem{Hartmann:2014fya}
  F.~Hartmann, W.~Kilian and K.~Schnitter,
  JHEP {\bf 1405} (2014) 064
  [arXiv:1401.7891].

\bibitem{King:2003rf}
  S.~F.~King and G.~G.~Ross,
  Phys.\ Lett.\ B {\bf 574} (2003) 239
  [hep-ph/0307190].
  
\bibitem{King:2006me}
  S.~F.~King and M.~Malinsky,
  JHEP {\bf 0611} (2006) 071
  [hep-ph/0608021].
 
\bibitem{King:2006np}
  S.~F.~King and M.~Malinsky,
  Phys.\ Lett.\ B {\bf 645} (2007) 351
  [hep-ph/0610250].
  
\bibitem{King:2009mk}
  S.~F.~King and C.~Luhn,
  Nucl.\ Phys.\ B {\bf 820} (2009) 269
  [arXiv:0905.1686].
 
\bibitem{Dutta:2009bj}
  B.~Dutta, Y.~Mimura and R.~N.~Mohapatra,
  JHEP {\bf 1005} (2010) 034
  [arXiv:0911.2242].
  
\bibitem{King:2009tj}
  S.~F.~King and C.~Luhn,
  Nucl.\ Phys.\ B {\bf 832} (2010) 414
  [arXiv:0912.1344].
  
\bibitem{Toorop:2010yh}
  R.~de Adelhart Toorop, F.~Bazzocchi and L.~Merlo,
  JHEP {\bf 1008} (2010) 001
  [arXiv:1003.4502].
 
\bibitem{BhupalDev:2011gi}
  P.~S.~Bhupal Dev, R.~N.~Mohapatra and M.~Severson,
  Phys.\ Rev.\ D {\bf 84} (2011) 053005
  [arXiv:1107.2378].

\bibitem{deMedeirosVarzielas:2011wx}
  I.~de Medeiros Varzielas,
  JHEP {\bf 1201} (2012) 097
  [arXiv:1111.3952].
 
\bibitem{BhupalDev:2012nm}
  P.~S.~Bhupal Dev, B.~Dutta, R.~N.~Mohapatra and M.~Severson,
  Phys.\ Rev.\ D {\bf 86} (2012) 035002
  [arXiv:1202.4012].

\bibitem{Hartmann:2014ppa}
  F.~Hartmann and W.~Kilian,
  Eur.\ Phys.\ J.\ C {\bf 74} (2014) 3055
  [arXiv:1405.1901].
  
\bibitem{King:2014iia}
  S.~F.~King,
  JHEP {\bf 1408} (2014) 130
  [arXiv:1406.7005].
  
\bibitem{Georgi:1979df}
  H.~Georgi and C.~Jarlskog,
  Phys.\ Lett.\ B {\bf 86} (1979) 297.

\bibitem{Beneke:2015lba}
  M.~Beneke, P.~Moch and J.~Rohrwild,
  Nucl.\ Phys.\ B {\bf 906} (2016) 561
  [arXiv:1508.01705].

\bibitem{Barr:1990vd} 
  S.~M.~Barr and A.~Zee,
  Phys.\ Rev.\ Lett.\  {\bf 65} (1990) 21
  [Phys.\ Rev.\ Lett.\  {\bf 65} (1990) 2920].
 
\bibitem{Chang:1993kw} 
  D.~Chang, W.~S.~Hou and W.~Y.~Keung,
  Phys.\ Rev.\ D {\bf 48} (1993) 217
  [hep-ph/9302267].
 
\bibitem{Kuno:1999jp} 
  Y.~Kuno and Y.~Okada,
  Rev.\ Mod.\ Phys.\  {\bf 73} (2001) 151
  [hep-ph/9909265].
 
\bibitem{Crivellin:2013ipa} 
  A.~Crivellin, M.~Hoferichter and M.~Procura,
  Phys.\ Rev.\ D {\bf 89} (2014) 054021 
  [arXiv:1312.4951].
  
\bibitem{Kitano:2002mt} 
  R.~Kitano, M.~Koike and Y.~Okada,
  Phys.\ Rev.\ D {\bf 66} (2002) 096002
  [Phys.\ Rev.\ D {\bf 76} (2007) 059902]
  [hep-ph/0203110].
  
\bibitem{Shifman:1978zn} 
  M.~A.~Shifman, A.~I.~Vainshtein and V.~I.~Zakharov,
  Phys.\ Lett.\ B {\bf 78} (1978) 443.

\bibitem{Crivellin:2014cta} 
  A.~Crivellin, M.~Hoferichter and M.~Procura,
  Phys.\ Rev.\ D {\bf 89} (2014) 093024
  [arXiv:1404.7134].

\bibitem{Aoki:2012zza} 
  M.~Aoki [DeeMe Collaboration],
  AIP Conf.\ Proc.\  {\bf 1441} (2012) 599.
 
\bibitem{Carey:2008zz} 
  R.~M.~Carey {\it et al.} [Mu2e Collaboration],
  FERMILAB-PROPOSAL-0973. 

\bibitem{Cui:2009zz} 
  Y.~G.~Cui {\it et al.} [COMET Collaboration],
  KEK-2009-10.  

\bibitem{TheMEG:2016wtm} 
  A.~M.~Baldini {\it et al.} [MEG Collaboration],
  arXiv:1605.05081.
  
\bibitem{Baldini:2013ke} 
  A.~M.~Baldini {\it et al.},
  arXiv:1301.7225.

\bibitem{Bellgardt:1987du} 
  U.~Bellgardt {\it et al.} [SINDRUM Collaboration],
  Nucl.\ Phys.\ B {\bf 299} (1988) 1.  
  
 \bibitem{Berger:2014vba} 
  N.~Berger [Mu3e Collaboration],
  Nucl.\ Phys.\ Proc.\ Suppl.\  {\bf 248-250} (2014) 35. 
  
  \bibitem{Bertl:2006up} 
  W.~H.~Bertl {\it et al.} [SINDRUM II Collaboration],
  Eur.\ Phys.\ J.\ C {\bf 47} (2006) 337.
  
\bibitem{Agashe:2006iy}
  K.~Agashe, A.~E.~Blechman and F.~Petriello,
  Phys.\ Rev.\ D {\bf 74} (2006) 053011
  [hep-ph/0606021].

\bibitem{Blanke:2007db}
M.~Blanke, A.~J.~Buras, B.~Duling, A.~Poschenrieder and C.~Tarantino,
  JHEP {\bf 0705 } (2007)  013
 [hep-ph/0702136].

\bibitem{otherLHT2}
  F.~del Aguila, J.~I.~Illana, M.~D.~Jenkins,
  JHEP {\bf 0901 } (2009)  080 
 [arXiv:0811.2891].

\bibitem{Blanke:2009am}
  M.~Blanke, A.~J.~Buras, B.~Duling, S.~Recksiegel, C.~Tarantino,
  Acta Phys.\ Polon.\  {\bf B41 } (2010)  657 
 [arXiv:0906.5454].

\bibitem{otherLHT1}
 T.~Goto, Y.~Okada, Y.~Yamamoto,
 Phys.\ Rev.\  {\bf D83 } (2011)  053011 
 [arXiv:1012.4385].

\bibitem{Ellis:2002fe}
J.~R.~Ellis, J.~Hisano, M.~Raidal and Y.~Shimizu,
  Phys.\ Rev.\  {\bf D66 } (2002)  115013 
  [hep-ph/0206110].

\bibitem{Brignole:2004ah}
  A.~Brignole, A.~Rossi,
  Nucl.\ Phys.\  {\bf B701 } (2004)  3
 [hep-ph/0404211].

\bibitem{Paradisi:2005tk}
 P.~Paradisi,
 JHEP {\bf 0602 } (2006)  050
 [hep-ph/0508054].

\bibitem{Paradisi:2006jp}
  P.~Paradisi,
  JHEP {\bf 0608 } (2006)  047 
 [hep-ph/0601100].

\bibitem{Calibbi:2009wk}
  L.~Calibbi, M.~Frigerio, S.~Lavignac and A.~Romanino,
  JHEP {\bf 0912} (2009) 057
  [arXiv:0910.0377]. 

\bibitem{Feruglio:2009hu}
F.~Feruglio, C.~Hagedorn, Y.~Lin and L.~Merlo,
  Nucl.\ Phys.\  {\bf B832 } (2010)  251 
  [arXiv:0911.3874].

\bibitem{Dimou:2015yng}
  M.~Dimou, S.~F.~King and C.~Luhn,
  JHEP {\bf 1602} (2016) 118
  [arXiv:1511.07886].

\bibitem{Dimou:2015cmw}
  M.~Dimou, S.~F.~King and C.~Luhn,
  Phys.\ Rev.\ D {\bf 93} (2016)  075026
  [arXiv:1512.09063].

\bibitem{Dev:2013oxa}
  C.~H.~Lee, P.~S.~Bhupal Dev and R.~N.~Mohapatra,
  Phys.\ Rev.\ D {\bf 88} (2013)  093010
  [arXiv:1309.0774]. 

\bibitem{Buras:2010cp}
A.~J.~Buras, B.~Duling, T.~Feldmann, T.~Heidsieck and C.~Promberger,
  JHEP {\bf 1009 } (2010)  104
 [arXiv:1006.5356].

\bibitem{Aubert:2009ag} 
  B.~Aubert {\it et al.} [BaBar Collaboration],
  Phys.\ Rev.\ Lett.\  {\bf 104} (2010) 021802 
  [arXiv:0908.2381].
  
\bibitem{Hayasaka:2010np}
  K.~Hayasaka {\it et al.},
  Phys.\ Lett.\ B {\bf 687} (2010) 139
  [arXiv:1001.3221].
  
\bibitem{Aaij:2013fia} 
  R.~Aaij {\it et al.} [LHCb Collaboration],
  Phys.\ Lett.\ B {\bf 724} (2013) 36 
  [arXiv:1304.4518].
  
\bibitem{Baron:2013eja} 
  J.~Baron {\it et al.} [ACME Collaboration],
  Science {\bf 343} (2014) 269
   [arXiv:1310.7534 ].
  
\bibitem{Davidson:2006bd}
  S.~Davidson and F.~Palorini,
  Phys.\ Lett.\ B {\bf 642} (2006) 72 
  [hep-ph/0607329].
  









  
\end{thebibliography}
\end{document}